\documentclass{iopjournal}
\usepackage[T1]{fontenc}
\usepackage[utf8]{inputenc}
\usepackage{lmodern,amsmath,amssymb,microtype,ragged2e,placeins}
\AtBeginDocument{\justifying}
\begin{document}
\title{Renormalization and exact Bloch bands: the trigonometric Rosen--Morse model}
\author{Ulysses Camara da Silva$^1$\orcid{0000-0002-0409-5284} and H. G. Vianna$^1$}

\affil{$^1$Department of Physics, Universidade Federal do Esp\'irito Santo, Vit\'oria, Brazil}
\email{ulysses.silva@ufes.br}
\begin{abstract}
\justifying
We study the trigonometric Rosen--Morse potential as an isolated singular well and as an exactly solvable periodic lattice. The potential combines inverse-square singularities with Coulomb terms of opposite signs. In the strongly attractive regime, independent renormalization at the two endpoints gives a two-parameter family of spectra. Deep negative-energy levels form two asymptotic sequences with the same Efimov-like scaling and opposite first-order Coulomb corrections. In the weak-medium regime, self-adjoint matching conditions allow transmission between neighboring cells. We obtain the exact Bloch dispersion relation and determine the band widths, gaps, and electron and hole effective masses. At fixed band index, the bandwidth becomes exponentially small in the limit of large Coulomb coupling.
\end{abstract}

\section{Introduction}
Inverse-square singularities give rise to quantum-mechanical problems
whose spectra depend on the boundary conditions imposed at the singular
points. Depending on the coupling, normalizability may admit two
independent local solutions and therefore fail to determine a unique
Hamiltonian \cite{Case1950,Frank1971}. Renormalization provides a physical
description of this ambiguity: the unresolved dynamics inside a small
core is encoded in boundary parameters that determine the solution outside it \cite{Renor_Inver_Squa,renor_orig,Burgess:2017ud}. The renormalized boundary parameters specify the domain of a self-adjoint Hamiltonian, providing an equivalent description of the freedom expressed by self-adjoint extensions \cite{Gitman}. These parameters can introduce a length scale into an interaction that is classically scale invariant \cite{deAlfaro:1976vlx,S_Coon}.

Near a singular point, chosen at $x=0$, we consider potentials
whose leading behavior is $2mV(x)/\hbar^2\approx g/x^2$,
where $g$ is the dimensionless inverse-square coupling.
The renormalized boundary conditions depend on its strength.
For $g\geq-1/4$, compatibility with supersymmetric quantum mechanics
selects the boundary condition that preserves the asymptotic
conformal symmetry \cite{Asim_paper,Lima:2019xzg}.
In the strong-attractive regime, $g<-1/4$, the local solutions
oscillate in the logarithm of the distance to the singularity.
Renormalization then introduces a boundary phase, and the running
coupling exhibits a limit cycle associated with discrete scaling
\cite{renor_orig,Bawin:2003dm,renor_Group_Limit_Cycle}.
Exactly solvable potentials with these singularities allow one to
follow how such local properties determine the complete spectrum.

This approach was applied to the generalized hyperbolic and
trigonometric P\"oschl--Teller potentials in
Refs.~\cite{CAMARADASILVA2024169549z,CamaraDaSilvaTPT}. The latter
model is defined between two inverse-square singularities and admits
independent renormalized boundary conditions at its endpoints.
A subsequent periodic construction connected neighboring singular
wells through a transmitting self-adjoint matching rule and yielded
exact Bloch bands \cite{CamaraDaSilvaLattice}. Such constructions
extend the study of singular Schr\"odinger operators from isolated
systems to periodic arrays
\cite{GesztesyMacedoStreit1985,GesztesyKirsch1985}.
The boundary conditions at the singular sites are essential: repeating
an isolated well with impenetrable endpoints does not by itself produce
a dispersive band structure.

In this work, we bring together these two analyses for the
trigonometric Rosen--Morse potential. This shape-invariant model is
exactly solvable \cite{dutt1988} and combines an inverse-square
singularity with a Coulomb term near each endpoint of its elementary
interval. The inverse-square coupling is the same at both ends,
whereas the Coulomb contribution has opposite signs. Although
subleading near either singularity, this additional interaction
changes the global spectrum and provides a solvable setting in which
to examine corrections to inverse-square scaling and their consequences
for the periodic system.

The general solution is expressed in terms of hypergeometric functions whose argument encircles a singular point in the complex plane as the coordinate runs across the cell. Accounting for the corresponding analytic continuation, we determine the relation between the asymptotic coefficients at the two endpoints. The resulting connection matrix provides the basis for studying both the renormalized isolated well and its periodic extension.
\begin{figure}[t]
\centering\includegraphics[scale=0.6]{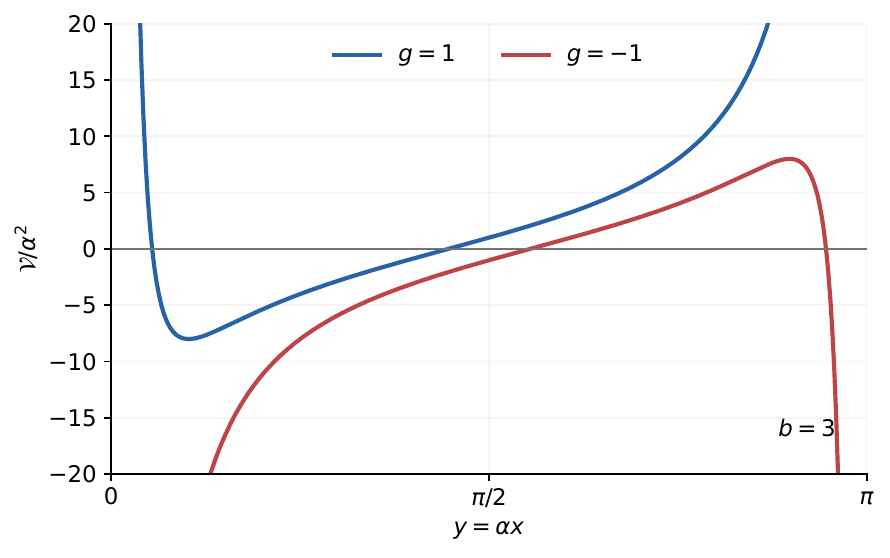}
\caption{The potential $\mathcal V/\alpha^2=g\csc^2y-2b\cot y$, with
$y=\alpha x$, for $g=1$ and $g=-1$, at $b=3$. The positive-$g$ curve has
an interior minimum, while the negative-$g$ curve has an interior maximum.
The vertical range is truncated near the singularities.}
\label{fig:potential}
\end{figure}

For the isolated cell, the SUSY-compatible boundary conditions recover
the known spectrum for $g\geq-1/4$. For $g<-1/4$, two independent
endpoint phases specify a family of spectra.
The deep negative-energy states organize into two asymptotic
sequences with the same inverse-square scaling factor and opposite
first Coulomb corrections. Their interlacing need not give a single
geometric sequence when all levels are ordered together. The ideal
singular Hamiltonian is unbounded from below; its effective
interpretation with a finite core applies only below the energy
scale at which that core is resolved.

We then construct the lattice in the range where both local modes are square-integrable, using the matching prescription of Ref.~\cite{CamaraDaSilvaLattice}. From the exact dispersion relation, we determine the band widths, gaps and electron and hole effective masses. We also calculate the Bloch-state probability densities and show that different inverse-square couplings can give the same band structure while changing the spatial distribution of the states within each cell.

The paper is organized as follows. Section~2 presents the general
solution and its endpoint connection. Section~3 summarizes the
coupling regimes and renormalized boundary conditions. Section~4
analyzes the isolated-cell spectrum and its deep-level scaling.
Section~5 develops the periodic extension, Bloch bands, effective
masses and probability densities. Section~6 presents our conclusions. Appendix~\ref{app:g0} treats the
logarithmic case $g=0$ of the periodic lattice.

\section{General solution and asymptotic behavior}
\begin{figure}[t]
\centering\includegraphics[width=.86\linewidth]{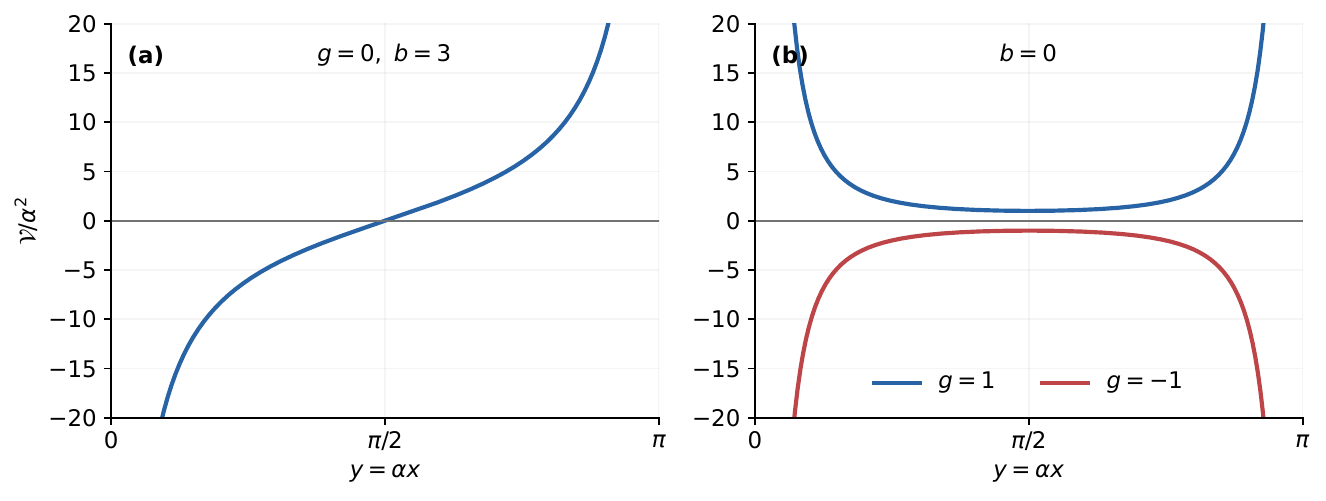}
\caption{Separate contributions to the potential: (a) $g=0$, $b=3$;
(b) $b=0$, with $g=1$ and $g=-1$. The Coulomb term has opposite signs at
the two ends, while the inverse-square term has the same sign at both ends.}
\label{fig:limits}
\end{figure}

Let us consider the trigonometric member of the Rosen--Morse family
\cite{RosenMorse1932}.\footnote{The original Rosen--Morse potential
is hyperbolic and regular. The trigonometric form also appears in
Schr\"odinger's treatment of the Coulomb problem on $S^3$
\cite{Schrodinger1940}, where its inverse-square coefficient is
$l(l+1)$, with $l$ the orbital angular-momentum quantum number.}
We define it on the elementary interval by
\begin{equation}
 \frac{2m}{\hbar^2}V(x)\equiv\mathcal V(x)
 =\alpha^2\left[\frac{g}{\sin^2(\alpha x)}-2b\cot(\alpha x)\right],
 \qquad 0<x<\frac\pi\alpha.
 \label{eq:potential}
\end{equation}
Here $\alpha>0$ has dimension of inverse length, while $g$ and $b$ are dimensionless.
We take $b\geq0$, since $\mathcal V(x;b)=\mathcal V(\pi/\alpha-x;-b)$: changing its sign
only interchanges the two singularities. For now, the model is defined in this
first cell; the periodic extension will be discussed later.
Figure~\ref{fig:potential} shows the potential for positive and negative $g$,
with the same Coulomb coupling. The separate contributions are shown in
Fig.~\ref{fig:limits}.

The stationary Schr\"odinger equation is
\begin{equation}
 \psi_k''(x)+[k^2-\mathcal V(x)]\psi_k(x)=0,
 \qquad k^2=\frac{2mE}{\hbar^2}.
\end{equation}
For negative energies, $k=i\kappa$, with $\kappa>0$. Near the singularities,
\begin{equation}
 \mathcal V(x)=\frac{g}{x^2}-\frac{2\alpha b}{x}+O(1),\quad x\to0;\qquad
 \mathcal V(x)=\frac{g}{(\pi/\alpha-x)^2}+\frac{2\alpha b}{\pi/\alpha-x}+O(1),\quad x\to\frac\pi\alpha.
\end{equation}
Thus the inverse-square coupling is the same at both ends, whereas the
Coulomb contribution is attractive at the left end and repulsive at the right end. Defining
\begin{equation}
\begin{gathered}
 \nu=\sqrt{\frac14+g},\qquad z(x)=1-e^{-2i\alpha x},\\[3pt]
 u(k)=\sqrt{\frac12\left(\sqrt{\frac{k^4}{\alpha^4}+4b^2}
                                  +\frac{k^2}{\alpha^2}\right)},\qquad
 \Sigma(u)=\frac12\left(u+\frac{ib}{u}\right),
 \qquad \frac{k^2}{\alpha^2}=u^2-\frac{b^2}{u^2},
\end{gathered}
 \label{eq:parameters}
\end{equation}
we choose $u(k)>0$ for $b>0$ and real $k^2$. The change $u\mapsto-u$
in the auxiliary functions gives $\Sigma(-u)=-\Sigma(u)$.
For $b=0$, expressions containing $b/u$ are understood as their
$b\to0^+$ limits at fixed $k^2$.
The parameter $\nu$ is real for $g\geq-1/4$ and becomes $\nu=i|\nu|$ for
$g<-1/4$. The general solution can be written directly in terms of $z$ as
\begin{equation}
 \psi_k(x)=A_0(k)\left(\frac z{2i\alpha}\right)^{1/2+\nu}
 \mathcal F(\nu;-u;z)
 +B_0(k)\left(\frac z{2i\alpha}\right)^{1/2-\nu}
 \mathcal F(-\nu;u;z),
 \label{eq:general}
\end{equation}
where
\begin{equation}
 \mathcal F(\nu;u;z)=(1-z)^{\Sigma(u)}
 {}_2F_1\left(\tfrac12+\nu+u,\tfrac12+\nu+ib/u;
 1+2\nu;z\right).
 \label{eq:F}
\end{equation}
Equation~\eqref{eq:general} is written for $2\nu\notin\mathbb Z$;
resonant values require limiting solutions \cite[\S15.10(i)]{NIST:DLMF}.
\subsection{Behavior at the first singularity}
The coefficients $A_0,B_0$ are normalized so that near $x=0$,
\begin{align}
 \psi_k(x)\approx{}&A_0(k)x^{1/2+\nu}[1+b\,O(x)+O(x^2)]\nonumber\\
 &+B_0(k)x^{1/2-\nu}[1+b\,O(x)+O(x^2)].
 \label{eq:left}
\end{align}
For real $\nu>1/2$, the Coulomb correction to the irregular branch is
larger than the leading term of the regular branch. This does not affect
the exact connection between the solutions derived below.

\subsection{Connection to $z=1$ and monodromy}
\begin{figure}[t]
\centering\includegraphics[scale=0.6]{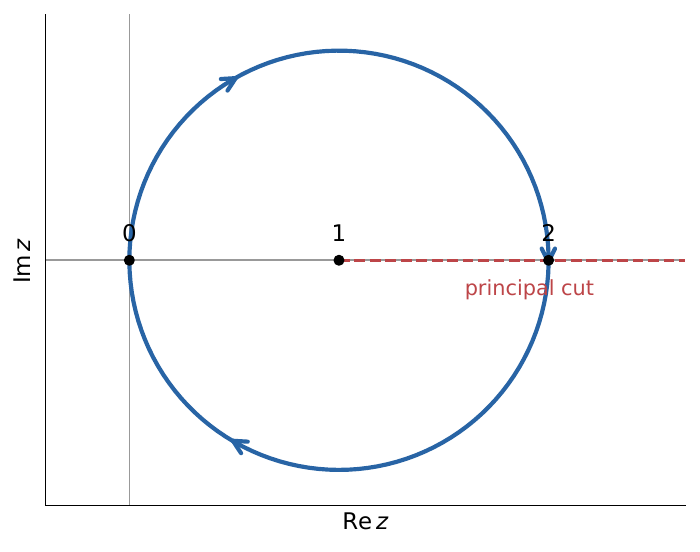}
\caption{The path $z=1-e^{-2iy}$, $0<y<\pi$, leaves the origin through
the upper half-plane, encircles $z=1$ clockwise, and returns through the
lower half-plane. The dashed ray denotes the principal cut $[1,\infty)$,
which is crossed at $z=2$, or $y=\pi/2$.}
\label{fig:path}
\end{figure}

The two endpoints both correspond to $z=0$, but the path between them
encircles $z=1$ clockwise, as shown in Fig.~\ref{fig:path}.
To follow this continuation, we first write the solution in a basis centered
at $z=1$, perform the rotation there, and then return to a basis centered at $z=0$.
The connection identity is \cite[Eq.~15.10.21]{NIST:DLMF}
\begin{align}
 {}_2F_1(a,b;c;z)={}&
 \frac{\Gamma(c)\Gamma(c-a-b)}{\Gamma(c-a)\Gamma(c-b)}
 {}_2F_1(a,b;a+b+1-c;1-z)\nonumber\\
 &+\frac{\Gamma(c)\Gamma(a+b-c)}{\Gamma(a)\Gamma(b)}
 (1-z)^{c-a-b}{}_2F_1(c-a,c-b;1+c-a-b;1-z).
 \label{eq:connection}
\end{align}

Define
\begin{equation}
 \zeta(\nu;u)=\frac{\Gamma(1+2\nu)\Gamma(2\Sigma(u))}
 {\Gamma(\tfrac12+\nu+u)\Gamma(\tfrac12+\nu+ib/u)},
 \label{eq:zeta}
\end{equation}
and the functions
\begin{equation}
 \mathcal G(\nu;u;1-z)=(1-z)^{\Sigma(u)}
 {}_2F_1\left(\tfrac12+\nu+u,\tfrac12+\nu+ib/u;
 1+2\Sigma(u);1-z\right).
\end{equation}
Applying Eq.~\eqref{eq:connection} to both terms of Eq.~\eqref{eq:general},
and using Euler's transformation to identify the same two complete solutions,
we obtain
\begin{align}
 \psi_k(x)={}&\left[\zeta(\nu;-u)A_0
 +\zeta(-\nu;-u)\frac{B_0}{(2i\alpha)^{-2\nu}}\right]
 \left(\frac z{2i\alpha}\right)^{1/2+\nu}\mathcal G(\nu;u;1-z)
 \nonumber\\
 &+\left[\zeta(\nu;u)\frac{A_0}{(2i\alpha)^{2\nu}}
 +\zeta(-\nu;u)B_0\right]
 \left(\frac z{2i\alpha}\right)^{1/2-\nu}\mathcal G(-\nu;-u;1-z).
 \label{eq:onebasis}
\end{align}
With the solution written in terms of hypergeometric functions centered
at $z=1$, we can follow the path from $x=0$ to $x=\pi/\alpha$ in
Fig.~\ref{fig:path}. The hypergeometric factors are analytic at $z=1$,
while the factors $(1-z)^{\pm\Sigma(u)}=e^{\mp2i\Sigma(u)\alpha x}$ acquire
$e^{\mp2\pi i\Sigma(u)}$ over this clockwise circuit. We then apply
Eq.~\eqref{eq:connection} again, now interchanging $z$ and $1-z$, to
express the continued solution in terms of hypergeometric functions
centered at $z=0$.

Since $z\approx-2i\alpha(\pi/\alpha-x)$ at the second singularity,
and the argument of $z/(2i\alpha)$ approaches $-\pi$ along the path,
\begin{equation}
 \left(\frac z{2i\alpha}\right)^{1/2\pm\nu}
 \approx e^{-i\pi(1/2\pm\nu)}
 \left(\frac\pi\alpha-x\right)^{1/2\pm\nu}.
\end{equation}
The resulting asymptotic form is
\begin{align}
 \psi_k(x)\approx{}&A_1(k)\left(\frac\pi\alpha-x\right)^{1/2+\nu}
 \left[1+b\,O\!\left(\frac\pi\alpha-x\right)
          +O\!\left(\left(\frac\pi\alpha-x\right)^2\right)\right]\nonumber\\
 &+B_1(k)\left(\frac\pi\alpha-x\right)^{1/2-\nu}
 \left[1+b\,O\!\left(\frac\pi\alpha-x\right)
          +O\!\left(\left(\frac\pi\alpha-x\right)^2\right)\right].
 \label{eq:right}
\end{align}
The coefficients are related by
\begin{equation}
 \begin{pmatrix}A_1/(2i\alpha)^\nu\\B_1/(2i\alpha)^{-\nu}\end{pmatrix}
 =M(k)\begin{pmatrix}A_0/(2i\alpha)^\nu\\B_0/(2i\alpha)^{-\nu}\end{pmatrix},\qquad
 M(k)=\begin{pmatrix}M_{11}&M_{12}\\M_{21}&M_{22}\end{pmatrix},
 \label{eq:matrix}
\end{equation}
where
\begin{align}
 M_{11}={}&\frac{\Sigma(u)}{\nu} e^{-i\pi(1/2+\nu)}
 \left[e^{2\pi i\Sigma(u)}\zeta(\nu;u)\zeta(-\nu;-u)
 -e^{-2\pi i\Sigma(u)}\zeta(\nu;-u)\zeta(-\nu;u)\right],\nonumber\\
 M_{22}={}&\frac{\Sigma(u)}{\nu} e^{-i\pi(1/2-\nu)}
 \left[e^{-2\pi i\Sigma(u)}\zeta(\nu;u)\zeta(-\nu;-u)
 -e^{2\pi i\Sigma(u)}\zeta(\nu;-u)\zeta(-\nu;u)\right],\nonumber\\
 M_{12}={}&\frac{2\Sigma(u)}{\nu}e^{-i\pi\nu}\sin(2\pi\Sigma(u))
 \zeta(-\nu;u)\zeta(-\nu;-u),\nonumber\\
 M_{21}={}&-\frac{2\Sigma(u)}{\nu}e^{i\pi\nu}\sin(2\pi\Sigma(u))
 \zeta(\nu;u)\zeta(\nu;-u).
 \label{eq:zetaentries}
\end{align}
Using the Gamma recurrence and reflection formulas
\cite[Eqs.~5.5.1 and 5.5.3]{NIST:DLMF},
\begin{equation}
 \Gamma(1+t)=t\Gamma(t),\qquad
 \Gamma(t)\Gamma(1-t)=\frac\pi{\sin(\pi t)},
\end{equation}
we obtain
\begin{align}
 M_{11}&=-\frac{\cos(\pi u)\cosh(\pi b/u)}{\sin(\pi\nu)}
 -\frac{\sin(\pi u)\sinh(\pi b/u)}{\cos(\pi\nu)},\nonumber\\
 M_{22}&=\frac{\cos(\pi u)\cosh(\pi b/u)}{\sin(\pi\nu)}
 -\frac{\sin(\pi u)\sinh(\pi b/u)}{\cos(\pi\nu)}.
\end{align}
The off-diagonal entries can also be written as
\begin{align}
M_{12}(k)={}&-\frac{\pi e^{-i\pi\nu}}
 {\nu\,\mathrm \mathrm B(\tfrac12-\nu+u,\tfrac12-\nu-u)\mathrm \mathrm B(\tfrac12-\nu+ib/u,\tfrac12-\nu-ib/u)},\nonumber\\
 M_{21}(k)={}&\frac{\pi e^{i\pi\nu}}
 {\nu\,\mathrm \mathrm B(\tfrac12+\nu+u,\tfrac12+\nu-u)
          \mathrm \mathrm B(\tfrac12+\nu+ib/u,\tfrac12+\nu-ib/u)},
 \label{eq:spec-beta}
\end{align}
where $\mathrm B(s,t)=\Gamma(s)\Gamma(t)/\Gamma(s+t)$ is the Euler beta function \cite[Eq.~5.12.1]{NIST:DLMF}. The matrix satisfies $\det M=-1$.

At $g=-1/4$ and $g=0$, the local solutions contain logarithmic terms.
Their asymptotic forms are given in Section~\ref{sec:renormalized-boundaries};
the nonlogarithmic basis used above is replaced by the corresponding
logarithmic basis when needed.

\section{Coupling regimes and renormalized boundary conditions}
\label{sec:renormalized-boundaries}
The boundary conditions for inverse-square singularities have been studied
extensively \cite{Case1950,Frank1971,Gitman}. Their renormalization fixes the
relative amplitudes of the two local solutions independently of the energy
\cite{renor_orig,Lima:2019xzg,CAMARADASILVA2024169549z,CamaraDaSilvaTPT}. For the isolated Rosen--Morse cell,
we impose separated boundary conditions at $x_0=0$ and $x_1=\pi/\alpha$.
Fixing the corresponding renormalized boundary parameters is equivalent
to choosing a self-adjoint extension of the singular Schr\"odinger
operator \cite{Gitman}. Thus, whenever both branches are square-integrable, the ratio $B_l(k)/A_l(k)$
is a boundary parameter, with $l=0,1$. Although the inverse-square coupling
$g=\nu^2-1/4$ is the same at both singularities, the two boundary conditions
are independent. The Coulomb term changes the subleading corrections,
but not this classification.

\subsection{Strong-repulsive coupling: $g\geq3/4$}
For $\nu\geq1$, only the branch with exponent $1/2+\nu$ is
square-integrable. The other branch has a divergent norm, logarithmically
so at $\nu=1$. Normalizability therefore requires
\begin{equation}
 B_0(k)=B_1(k)=0.
 \label{eq:bc-repulsive}
\end{equation}
There is no boundary parameter to be fixed in this regime.

\subsection{Weak-medium coupling: $-1/4\leq g<3/4$}
For $0<\nu<1$, both local branches are square-integrable. Self-adjointness
requires a real, energy-independent ratio $B_l/A_l$ at each end
\cite{renor_m_w_c}. Generic
renormalized boundary conditions introduce a length scale; the choice
$B_l=0$ introduces no anomalous boundary scale and preserves the asymptotic
conformal symmetry \cite{Lima:2019xzg,CAMARADASILVA2024169549z,CamaraDaSilvaTPT}.

At $g=-1/4$ ($\nu=0$), the two exponents coincide and the local solution becomes
\begin{equation}
 \psi_k(x)\approx\sqrt{|x-x_l|}
 \left[A_l(k)+B_l(k)\ln\bigl(\alpha|x-x_l|\bigr)\right],
 \qquad l=0,1.
 \label{eq:bc-critical}
\end{equation}
Both terms remain square-integrable. Renormalization again fixes an
energy-independent relation between $A_l$ and $B_l$, with a boundary
scale when $B_l\neq0$. This is the critical coupling at which the two
real renormalization-group fixed points merge.

At $g=0$ ($\nu=1/2$), the exponents are $1$ and $0$, and the Coulomb
term introduces a logarithmic correction to the second local solution.
Writing $\xi=|x-x_l|$, with the same logarithmic reference scale at
both endpoints, gives
\begin{align}
 \psi_k(x)={}&A_l(k)\,\xi[1+O(\xi)]\nonumber\\
 &+B_l(k)\left[1-2(-1)^l\alpha b\,\xi\ln(\alpha\xi)
 +O(\xi^2|\ln(\alpha\xi)|)\right],\qquad l=0,1.
 \label{eq:bc-coulomb-log}
\end{align}
Both solutions are square-integrable.
The correction $\xi\ln(\alpha\xi)$ vanishes at the endpoint, whereas
its derivative diverges logarithmically. 
Requiring compatibility with SUSY selects $B_l=0$,
eliminating the logarithmic terms at both $g=0$ ($\nu=1/2$)
and $g=-1/4$ ($\nu=0$) \cite{Asim_paper,Lima:2019xzg}. We adopt this prescription
for the isolated cell. The periodic realization of the logarithmic
basis in Eq.~\eqref{eq:bc-coulomb-log} is derived in
Appendix~\ref{app:g0}.

\subsection{Strong-attractive coupling: $g<-1/4$}
For $\nu=i|\nu|$, the two square-integrable local solutions are complex
conjugates and oscillate in $\ln|x-x_l|$. Choosing real eigenfunctions gives
\begin{equation}
 B_l(k)=A_l^*(k), \ l=0,1.
 \label{eq:bc-reality}
\end{equation}
The renormalized boundary conditions can then be written as \cite{CAMARADASILVA2024169549z}
\begin{equation}
 (2\alpha)^{2i|\nu|}\frac{B_l(k)}{A_l(k)}=e^{i\theta_l},
 \label{eq:bc-phases}
\end{equation}
where $\theta_0$ and $\theta_1$ are independent of the energy and defined
modulo $2\pi$. We choose $2\alpha$ as the reference scale for convenience. The two phases specify a two-parameter family of self-adjoint
Hamiltonians. The associated short-distance running is periodic in the logarithm of the cutoff, with period $\pi/|\nu|$ \cite{CAMARADASILVA2024169549z,CamaraDaSilvaTPT}. The ideal singular Hamiltonian is not bounded from below. The spectra selected by these conditions and by $B_l=0$ in the preceding regimes will be discussed in the next section.
\section{Energy spectrum of the isolated cell}
\label{sec:cell-spectrum}
\subsection{The SUSY-selected boundary conditions: $\nu\geq0$}
For real $\nu\geq0$, compatibility with SUSY protects the asymptotic
conformal symmetry by selecting $B_l(k)=0$
\cite{Asim_paper,Lima:2019xzg}. For $\nu\geq1$, the same condition
already follows from normalizability. A nontrivial solution with $B_0=B_1=0$ requires
\begin{equation}
 M_{21}(k)=0.
 \label{eq:spec-regular-condition}
\end{equation}
For real $\nu>0$ and real energies, the relevant poles of the beta function in Eq.~\eqref{eq:spec-beta} occur at
\begin{equation}
 \tfrac12+\nu-u=-n,\qquad n=0,1,2,\ldots .
 \label{eq:spec-poles}
\end{equation}
Using $k^2/\alpha^2=u^2-b^2/u^2$, we obtain
\begin{equation}
 E_n=\frac{\hbar^2\alpha^2}{2m}
 \left[(n+1/2+\nu)^2-\frac{b^2}{(n+1/2+\nu)^2}\right],
 \qquad \nu\geq0.
 \label{eq:spec-regular}
\end{equation}
For $b\gg (1/2+\nu)^2$, the lowest modes with $n+1/2+\nu\ll\sqrt b$ are dominated
by the attractive Coulomb contribution, $E_n\approx
-\hbar^2\alpha^2b^2/[2m(n+1/2+\nu)^2]$. Conversely, for $n+1/2+\nu\gg\sqrt b$,
the spectrum is dominated by the quadratic term associated with
confinement to the cell. The inverse-square singularities shift the
level index by $1/2+\nu$, as in the trigonometric P\"oschl--Teller model
\cite{CamaraDaSilvaTPT}. Thus the same exact spectrum displays a Coulomb-like
low-energy sector and the large-$n$ behavior of a box with singular ends.

\begin{figure}[!htbp]
\centering\includegraphics[width=.98\linewidth]{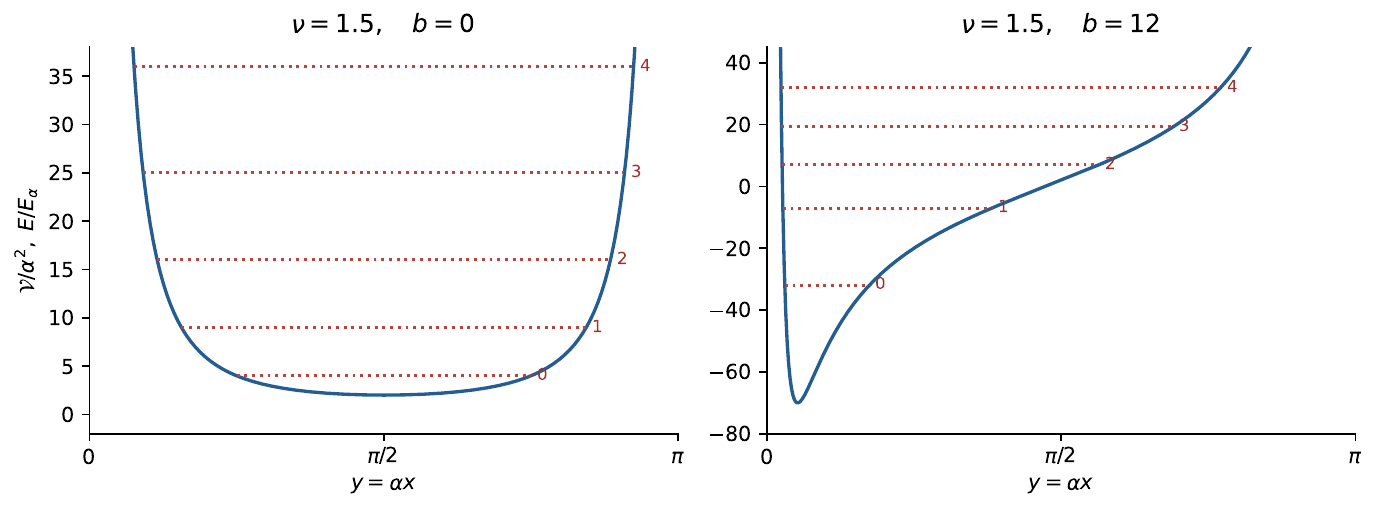}
\caption{The potential and the first five levels for $\nu=1.5$, with $b=0$
(left) and $b=12$ (right). Energies are measured in
$E_\alpha=\hbar^2\alpha^2/(2m)$. Dotted segments indicate the classically
allowed part of each level and are labeled by $n$. The vertical range is
truncated near the singularities.}
\label{fig:regular-levels}
\end{figure}
Figure~\ref{fig:regular-levels} illustrates these two contributions.
The same spectrum follows from the algebraic factorization of the
shape-invariant Hamiltonian \cite{dutt1988}.

\subsection{Strong-attractive coupling: $\nu=i|\nu|$}
We now impose the two independent boundary phases in Eq.~\eqref{eq:bc-phases}. The connection matrix in Eq.~\eqref{eq:matrix} relates the coefficient ratios at the two ends:
\begin{equation}
 (2i\alpha)^{2\nu}\frac{B_1}{A_1}
 =\frac{M_{21}+M_{22}(2i\alpha)^{2\nu}(B_0/A_0)}
 {M_{11}+M_{12}(2i\alpha)^{2\nu}(B_0/A_0)},\qquad \nu\to i|\nu|.
\end{equation}
Substituting Eq.~\eqref{eq:bc-phases} and clearing the denominator gives the spectral condition
\begin{equation}
 e^{\pi|\nu|}M_{21}-e^{-\pi|\nu|}e^{i(\theta_0+\theta_1)}M_{12}
 +e^{i\theta_0}M_{22}-e^{i\theta_1}M_{11}=0.
 \label{eq:spec-complex}
\end{equation}
For real energies, this condition reduces to a single real equation. Introducing the half-sum and half-difference
\begin{equation}
 \theta_+=\frac{\theta_0+\theta_1}{2},\qquad
 \theta_-=\frac{\theta_0-\theta_1}{2},
\end{equation}
and writing the relevant matrix entries explicitly, we obtain
\begin{align}
 e^{\pi|\nu|}M_{21}(k)&=e^{-\pi|\nu|}M_{12}^*(k)\nonumber\\
 &=\frac{\pi}
 {i|\nu|\,\mathrm B(\tfrac12+i|\nu|+u,\tfrac12+i|\nu|-u)
 \mathrm B(\tfrac12+i|\nu|+ib/u,\tfrac12+i|\nu|-ib/u)},
 \label{eq:spec-K}
\end{align}
\begin{equation}
 M_{11}=M_{22}^*
 =i\frac{\cos(\pi u)\cosh(\pi b/u)}{\sinh(\pi|\nu|)}
 -\frac{\sin(\pi u)\sinh(\pi b/u)}{\cosh(\pi|\nu|)}.
\end{equation}
Dividing Eq.~\eqref{eq:spec-complex} by $2i e^{i\theta_+}$ gives the
condition determining the energy spectrum,
\begin{equation}
 \operatorname{Im}\!\left[e^{-i\theta_+}e^{\pi|\nu|}M_{21}(k)\right]
 =\frac{\cos(\pi u)\cosh(\pi b/u)}{\sinh(\pi|\nu|)}\cos\theta_-
 +\frac{\sin(\pi u)\sinh(\pi b/u)}{\cosh(\pi|\nu|)}\sin\theta_-.
 \label{eq:spec-real}
\end{equation}
Thus the boundary phases enter through their sum and difference, with
the energy dependence contained in $u(k)$ and $M_{21}(k)$. For $b\neq0$, both terms on the right are generally
present. For $b=0$, the term proportional to
$\sin\theta_-$ vanishes; the equation is then invariant under interchange
of the two boundary phases, as required by reflection symmetry.
The real roots of Eq.~\eqref{eq:spec-real} determine the energies for
each pair $(\theta_+,\theta_-)$. Thus the two boundary angles specify a family of energy spectra and corresponding eigenstates. The phase coordinates inherit the identifications $(\theta_+,\theta_-)\equiv(\theta_++\pi,\theta_-\pm\pi)$ from the $2\pi$ periodicity of each endpoint phase.

\begin{figure}[t]
\centering\includegraphics[width=.98\linewidth]{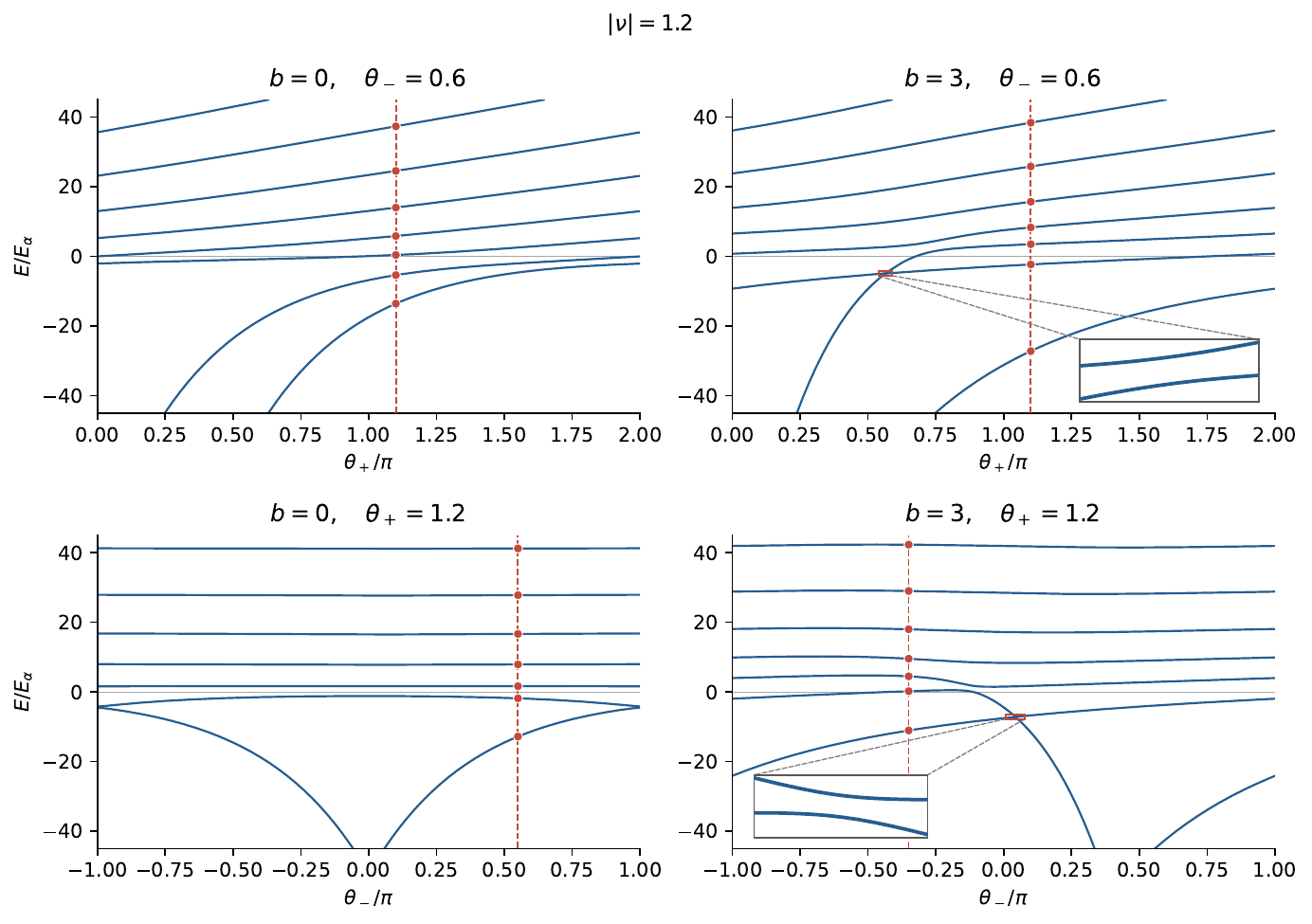}
\caption{Energy levels for $|\nu|=1.2$, at $b=0$ (left) and $b=3$
(right). Upper row: $\theta_+$ varies at fixed $\theta_-=0.6$; lower row:
$\theta_-$ varies at fixed $\theta_+=1.2$. The red vertical lines select
$\theta_+=1.1\pi$ (upper row), $\theta_-=0.55\pi$ (lower left), and
$\theta_-=-0.35\pi$ (lower right); the red dots mark the corresponding
energy levels. The insets show that the close branches do not cross.}
\label{fig:strong-flow}
\end{figure}
\begin{figure}[t]
\centering\includegraphics[width=.98\linewidth]{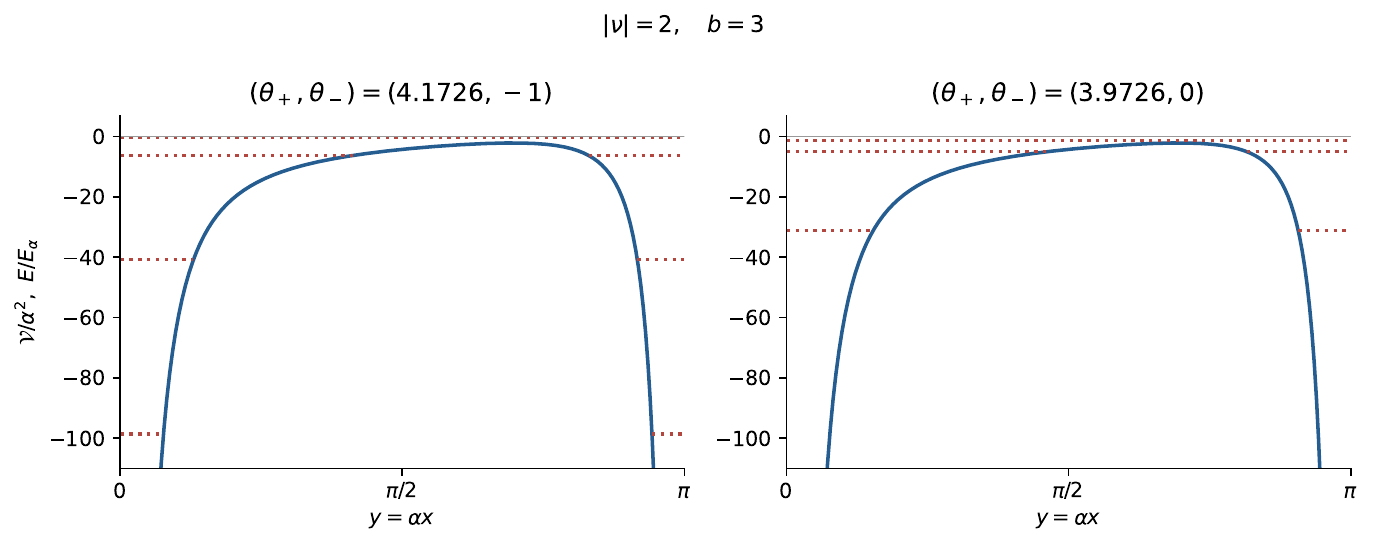}
\caption{The strong-attractive potential and all negative-energy levels
in the window $-110<E/E_\alpha<0$, for $|\nu|=2$ and $b=3$.
The same potential is shown for two choices of the boundary phases,
$(\theta_+,\theta_-)=(1.4+4\ln2,-1)$ (left) and $(1.2+4\ln2,0)$ (right), in radians. Dotted segments mark the classically allowed parts
of each energy level.
The levels are roots of the exact finite-cell spectral equation.
There are infinitely many levels below the displayed window.}
\label{fig:strong-levels}
\end{figure}

Figure~\ref{fig:strong-flow} shows how both positive and negative levels
depend on the boundary phases. Each vertical line selects the spectrum
for a fixed pair $(\theta_-,\theta_+)$; the red dashed lines and dots give
one example in each panel. 
Figure~\ref{fig:strong-levels} shows the energy levels directly on the potential curve. The localization of sufficiently deep states will be used in the following analysis.

\subsubsection{Deep levels and the Efimov limit}
For $E=-\hbar^2\kappa^2/(2m)$, with $\kappa>0$, the probability
density of sufficiently deep states is concentrated near the
endpoints of the cell. When $b=0$ and $\theta_0=\theta_1$, reflection
symmetry gives parity eigenstates with equal probability in the two
halves of the cell. To determine the contribution of each endpoint
to the deep spectrum, we write $\xi=|x-x_l|$ and expand the
potential near $x_l$,
\begin{equation}
 \mathcal V(x)=\frac{g}{\xi^2}-\frac{2(-1)^l\alpha b}{\xi}+O(\alpha^2),
 \qquad l=0,1.
\end{equation}
On the scale $\xi\approx\kappa^{-1}$, the Coulomb term gives a correction
of order $\varepsilon=\alpha b/\kappa$, whereas the bounded terms enter
at order $\alpha^2/\kappa^2$. We first solve the auxiliary half-line
problem retaining the two singular terms. Its Schr\"odinger equation is
\begin{equation}
 \psi_l''(\xi)+\left[-\kappa^2+\frac{2(-1)^l\alpha b}{\xi}
 +\frac{1/4+|\nu|^2}{\xi^2}\right]\psi_l(\xi)=0,
 \qquad \varepsilon=\frac{\alpha b}{\kappa}.
\end{equation}
The two solutions normalized at the singularity give
\begin{align}
 \psi_l(\xi)={}&A_l\xi^{1/2+i|\nu|}e^{-\kappa\xi}
 {}_1F_1\left(\tfrac12+i|\nu|-(-1)^l\varepsilon;
 1+2i|\nu|;2\kappa\xi\right)\nonumber\\
 &+B_l\xi^{1/2-i|\nu|}e^{-\kappa\xi}
 {}_1F_1\left(\tfrac12-i|\nu|-(-1)^l\varepsilon;
 1-2i|\nu|;2\kappa\xi\right).
 \label{eq:confluent-local}
\end{align}
Since ${}_1F_1(a;c;0)=1$, the coefficients $A_l,B_l$ coincide with
those used in the boundary conditions.

For large argument, the two contributions to the confluent
hypergeometric asymptotic expansion are \cite[Eq.~13.7.2]{NIST:DLMF}
\begin{equation}
 {}_1F_1(a;c;t)\approx
 \frac{\Gamma(c)}{\Gamma(a)}e^t t^{a-c}[1+O(t^{-1})]
 +\frac{\Gamma(c)}{\Gamma(c-a)}(-t)^{-a}[1+O(t^{-1})].
\end{equation}
The power $(-t)^{-a}$ is defined by lateral continuation from the upper or lower half-plane. The growing exponential has the same coefficient in either continuation. In Eq.~\eqref{eq:confluent-local}, the growing contribution is
\begin{align}
 \psi_l(\xi)\big|_{\mathrm{grow}}\approx{}&
 e^{\kappa\xi}(2\kappa\xi)^{-(-1)^l\varepsilon}
 \Bigg[
 A_l(2\kappa)^{-1/2-i|\nu|}
 \frac{\Gamma(1+2i|\nu|)}
 {\Gamma(\tfrac12+i|\nu|-(-1)^l\varepsilon)}\nonumber\\
 &\hspace{17mm}+B_l(2\kappa)^{-1/2+i|\nu|}
 \frac{\Gamma(1-2i|\nu|)}
 {\Gamma(\tfrac12-i|\nu|-(-1)^l\varepsilon)}
 \Bigg]\,[1+O((\kappa\xi)^{-1})].
\end{align}
Normalizability of the auxiliary half-line solution eliminates this
exponential. Setting its coefficient to zero and using the Gamma
recurrence yields
\begin{equation}
 \frac{B_l}{A_l}=(2\kappa)^{-2i|\nu|}
 \frac{\Gamma(2i|\nu|)}{\Gamma(-2i|\nu|)}
 \frac{\Gamma(\tfrac12-i|\nu|-(-1)^l\varepsilon)}
 {\Gamma(\tfrac12+i|\nu|-(-1)^l\varepsilon)}.
\end{equation}
Comparing with $(2\alpha)^{2i|\nu|}B_l/A_l=e^{i\theta_l}$ gives
\begin{equation}
 e^{i\theta_l}\approx
 \left(\frac{\alpha}{\kappa}\right)^{2i|\nu|}
 \frac{\Gamma(2i|\nu|)}{\Gamma(-2i|\nu|)}
 \frac{\Gamma(\tfrac12-i|\nu|-(-1)^l\varepsilon)}
 {\Gamma(\tfrac12+i|\nu|-(-1)^l\varepsilon)}.
 \label{eq:confluent-quantization}
\end{equation}
This approximation gives the spectrum of the deep levels associated
with each endpoint. The first-order Coulomb correction has opposite
signs in the two sequences, reflecting the attractive interaction at
one end of the cell and the repulsive interaction at the other.

At leading order, $\varepsilon=0$, the Gamma duplication identity gives
\begin{equation}
 e^{i\theta_l}\approx
 \left(\frac{4\alpha}{\kappa}\right)^{2i|\nu|}
 \frac{\Gamma(i|\nu|)}{\Gamma(-i|\nu|)}.
\end{equation}
Each endpoint therefore generates the asymptotic sequence
\begin{equation}
 \kappa^{(0)}_{l,n}=4\alpha
 \exp\left[\frac{\arg\Gamma(i|\nu|)-\theta_l/2+\pi n}{|\nu|}\right],
 \qquad n\to+\infty.
 \label{eq:efimov-sequences}
\end{equation}
Changing the branch of $\arg\Gamma(i|\nu|)$ relabels the integers.
The discrete scaling is
\begin{equation}
 \frac{\kappa_{l,n+1}}{\kappa_{l,n}}\longrightarrow e^{\pi/|\nu|},
 \qquad
 \frac{|E_{l,n+1}|}{|E_{l,n}|}\longrightarrow e^{2\pi/|\nu|},
 \label{eq:efimov-ratios}
\end{equation}
as for the inverse-square interaction
\cite{Bawin:2003dm,renor_Group_Limit_Cycle,CamaraDaSilvaTPT}.
This geometric scaling is analogous to the scaling of Efimov's
three-body states \cite{Efimov}; here it describes deep states
localized near the endpoints of a one-dimensional cell.

To retain the first Coulomb correction, expand the logarithm of
Eq.~\eqref{eq:confluent-quantization} in $\varepsilon$:
\begin{equation}
 |\nu|\ln\frac{\kappa}{4\alpha}
 =\arg\Gamma(i|\nu|)-\frac{\theta_l}{2}+\pi n
 +(-1)^l\varepsilon\operatorname{Im}\psi(\tfrac12+i|\nu|)
 +O\!\left(\frac{\alpha^2(1+b^2)}{\kappa^2}\right).
\end{equation}
Here $\psi$ is the digamma function. Its reflection identity
\cite[Eq.~5.5.4]{NIST:DLMF} gives
$\operatorname{Im}\psi(\tfrac12+i|\nu|)=(\pi/2)\tanh(\pi|\nu|)$.
Thus, for fixed $|\nu|$ and $b$,
\begin{equation}
 \kappa_{l,n}=\kappa^{(0)}_{l,n}
 +(-1)^l\frac{\pi\alpha b}{2|\nu|}\tanh(\pi|\nu|)
 +O\!\left(\frac{\alpha^2(1+b^2)}{\kappa^{(0)}_{l,n}}\right).
 \label{eq:coulomb-deep-shift}
\end{equation}
The attractive Coulomb term increases $\kappa$ at $x_0$; the repulsive
term decreases it at $x_1$. Consequently,
\begin{align}
 \frac{\kappa_{l,n+1}}{\kappa_{l,n}}={}&e^{\pi/|\nu|}
 -(-1)^l\bigl(e^{\pi/|\nu|}-1\bigr)
 \frac{\pi\alpha b\tanh(\pi|\nu|)}{2|\nu|\kappa^{(0)}_{l,n}}
 +O\!\left(\frac{\alpha^2(1+b^2)}{(\kappa^{(0)}_{l,n})^2}\right).
 \label{eq:ratio-coulomb}
\end{align}
The two sequences approach the same limit from opposite sides. The
first corrections decrease with the depth of the level.

\begin{table}[t]
\centering
\caption{Convergence of the ratios $R_{l,n}=\kappa_{l,n+1}/\kappa_{l,n}$
for the same parameters as Fig.~\ref{fig:efimov}: $|\nu|=2$, $b=3$ and
$(\theta_0,\theta_1)=(0.4+4\ln2,2.4+4\ln2)$ radians. Values are obtained directly
from roots of the exact finite-cell spectral equation. The index $n$
labels the two sequences according to Eq.~\eqref{eq:efimov-sequences},
not the ordering of the complete spectrum. The last two columns give
$\delta_{l,n}=|R_{l,n}/e^{\pi/2}-1|$, the relative deviation from the limiting ratio.}
\label{tab:efimov-ratios}
\begin{tabular}{crrrr}
\hline
 $n$ & $R_{0,n}$ & $R_{1,n}$ & $\delta_{0,n}$ & $\delta_{1,n}$\\
\hline
 1 & 3.551483 & 3.969018 & 0.261719 & 0.174922 \\
 2 & 4.423770 & 5.748006 & 0.080389 & 0.194893 \\
 3 & 4.721434 & 4.969042 & 0.018510 & 0.032962 \\
 4 & 4.791546 & 4.842190 & 0.003935 & 0.006592 \\
 5 & 4.806523 & 4.817018 & 0.000822 & 0.001360 \\
\hline
 Limit & 4.810477 & 4.810477 & 0 & 0\\
\hline
\end{tabular}
\end{table}
Table~\ref{tab:efimov-ratios} quantifies the convergence to
$e^{\pi/2}=4.810477381\ldots$. In the deep regime, the attractive
Coulomb family approaches this limit from below and the repulsive
family from above, consistently with Eq.~\eqref{eq:coulomb-deep-shift}.
\begin{figure}[t]
\centering\includegraphics[scale=0.65]{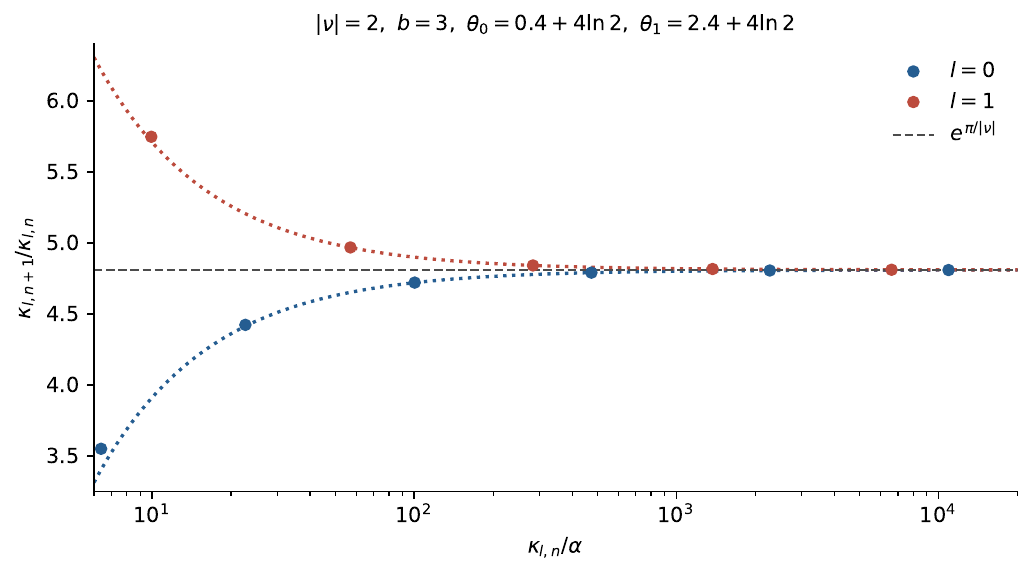}
\caption{Ratios of successive deep levels within each endpoint family,
for $|\nu|=2$, $b=3$, $\theta_0=0.4+4\ln2$ and
$\theta_1=2.4+4\ln2$. The dashed horizontal line is $e^{\pi/|\nu|}$.
The dots mark the exact levels; the colored dotted curves include the
first Coulomb correction for each singularity. The attractive family approaches the limit from below and
the repulsive family from above.}
\label{fig:efimov}
\end{figure}
The exact roots confirm the opposite approaches in Fig.~\ref{fig:efimov}.

The absence of a lower bound is a property of the ideal singular
Hamiltonian. Its renormalization has an effective interpretation:
the potential inside each core $\xi<R$ is replaced by unknown
short-distance physics, whose leading effect is encoded in
$\theta_0$ and $\theta_1$ \cite{CAMARADASILVA2024169549z}.
The energy-independent boundary conditions assume $\kappa R\ll1$,
besides $\alpha R\ll1$. Thus the deep-level expansion has a physical
window in which
\begin{equation}
 \frac{\alpha}{\kappa}\ll1,\qquad
 \frac{\alpha b}{\kappa}\ll1,\qquad
 \kappa R\ll1.
 \label{eq:deep-validity}
\end{equation}
At fixed $R$, only a finite part of the deep negative-energy tower
lies in this window. When $\kappa\sim1/R$, the core is resolved and
the energy-independent phases no longer suffice. A nonsingular core
that makes the full potential bounded from below restores a ground
state, whose energy cannot be inferred from these phases alone.
The infinite descent and the $n\to\infty$ scaling refer to the ideal
zero-cutoff model.

\begin{figure}[t]
\centering\includegraphics[scale=0.6]{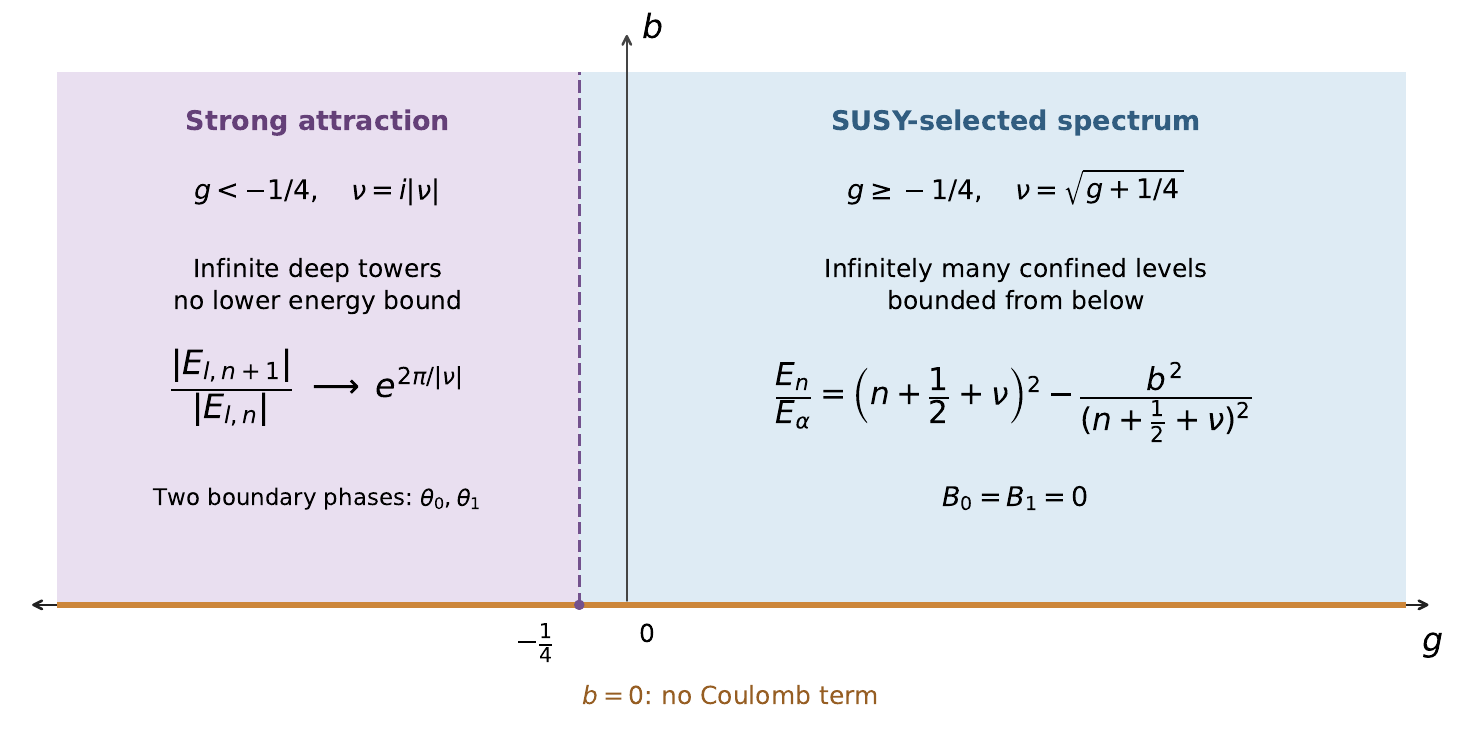}
\caption{Isolated-cell spectra in the $(g,b)$ plane, for the boundary
conditions adopted in this work. The vertical line marks $g=-1/4$;
the highlighted horizontal line is the Coulomb-free limit $b=0$.
The energy ratio on the left refers to successive deep levels within
each endpoint family of the ideal singular model.}
\label{fig:spectral-map}
\end{figure}

Figure~\ref{fig:spectral-map} summarizes the isolated-cell spectra for
the boundary conditions adopted here. For $g\geq-1/4$, the
SUSY-selected spectrum is bounded from below and has infinitely many
confined levels, with the exact quadratic and Coulomb contributions
in Eq.~\eqref{eq:spec-regular}. For $g<-1/4$, two endpoint phases
specify the spectrum and its two deep Efimov sequences. The line
$b=0$ is the Coulomb-free limit: the regular spectrum becomes purely
quadratic and the opposite first-order shifts in
Eq.~\eqref{eq:coulomb-deep-shift} vanish. The critical inverse-square
coupling remains $g=-1/4$.

\section{Periodic extension and Bloch bands}
\label{sec:lattice}
\begin{figure}[t]
\centering\includegraphics[width=.98\linewidth]{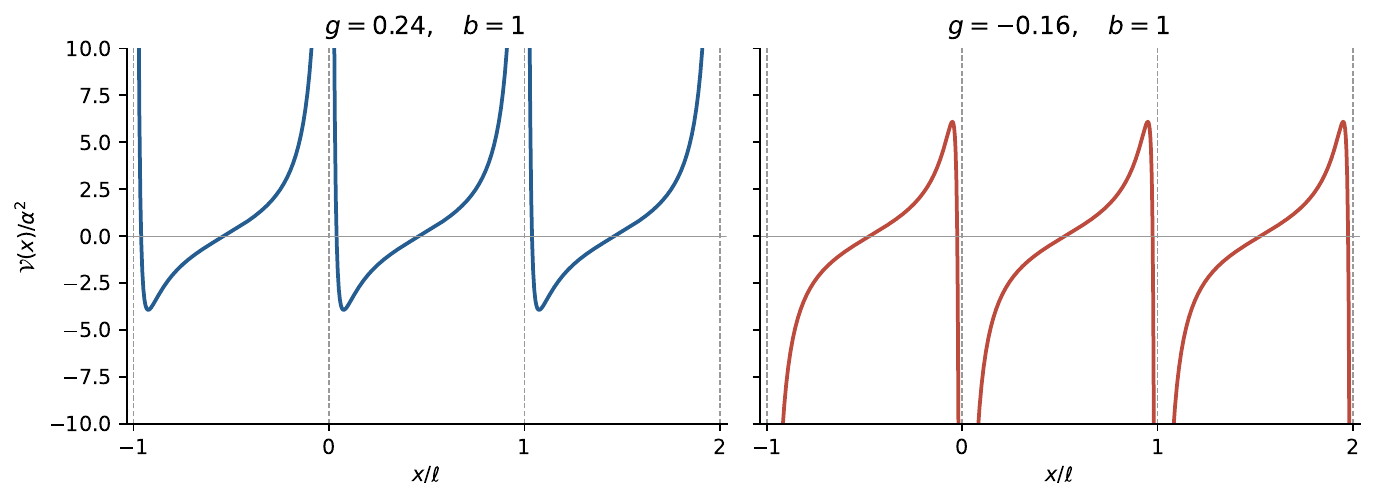}
\caption{Three cells of the periodic Rosen--Morse potential, for $b=1$,
with $g=0.24$ (left) and $g=-0.16$ (right). Dashed vertical lines mark
the singularities; the vertical range is truncated.}
\label{fig:periodic-potential}
\end{figure}

We now extend the model to the real line, following the construction
of the trigonometric P\"oschl--Teller lattice \cite{CamaraDaSilvaLattice}.
The potential in Eq.~\eqref{eq:potential} is continued periodically,
with period $\ell=\pi/\alpha$ and singularities at $x_j=j\pi/\alpha$,
$j\in\mathbb Z$. Using the partial-fraction identities in
Ref.~\cite[Eqs.~4.22.3--4.22.4]{NIST:DLMF}, the potential can be
rewritten as
\begin{equation}
 \mathcal V(x)=g\sum_{j\in\mathbb Z}\frac{1}{(x-j\ell)^2}
 -2\alpha b\left\{\frac1x+\sum_{j=1}^{\infty}
 \left[\frac{1}{x-j\ell}+\frac{1}{x+j\ell}\right]\right\}.
 \label{eq:periodic-partial-fractions}
\end{equation}
The sign of the Coulomb contribution changes across each singularity,
while the inverse-square contribution has the same sign on both sides.
Figure~\ref{fig:periodic-potential} shows three cells
for positive and negative $g$. Its realization on the line requires
boundary conditions connecting the solutions across every singularity.

Throughout this section we restrict the coupling to
\begin{equation}
 0<\nu<1,\qquad \nu\neq\tfrac12,
\end{equation}
so that both local modes are square-integrable. The logarithmic case
$\nu=1/2$ is treated in Appendix~\ref{app:g0}, where the same dispersion
relation is recovered from finite endpoint coefficients. The conditions
used for the isolated cell are replaced by transmitting matching
conditions.

\subsection{Matching across the singularities}
Near a generic $x_j$, let $A_<,B_<$ and $A_>,B_>$ denote the coefficients
on its left and right, respectively. For $\nu\neq1/2$, in the normalization already used,
\begin{align}
 \psi_k(x)\approx{}&
 A_{\lessgtr}(k)|x-x_j|^{1/2+\nu}
       [1+b\,O(|x-x_j|)+O(|x-x_j|^2)]\nonumber\\
 &+B_{\lessgtr}(k)|x-x_j|^{1/2-\nu}
       [1+b\,O(|x-x_j|)+O(|x-x_j|^2)].
 \label{eq:lattice-local}
\end{align}
The coefficients refer to the full local solutions, including the
Coulomb corrections, whose linear terms have opposite signs on the
two sides. Following Ref.~\cite{CamaraDaSilvaLattice}, consider the
boundary form associated with two functions in the Hamiltonian domain,
\begin{equation}
 J_{k'k}(x)=\psi_{k'}^*(x)\frac{d\psi_k(x)}{dx}
 -\frac{d\psi_{k'}^*(x)}{dx}\psi_k(x).
 \label{eq:boundary-current}
\end{equation}
Integration by parts on the intervals on either side of $x_j$ gives
an internal boundary contribution proportional to
$J_{k'k}(x_j^-)-J_{k'k}(x_j^+)$. For this contribution to vanish,
the matching must satisfy
\begin{equation}
 J_{k'k}(x_j^-)=J_{k'k}(x_j^+)
 \label{eq:boundary-continuity}
\end{equation}
for every pair of functions in the domain. In particular,
$\hbar J_{kk}/(2mi)$ is the probability current. Substitution of
Eq.~\eqref{eq:lattice-local}, with the opposite orientations of
$|x-x_j|$ on the two sides, gives
\begin{align}
 J_{k'k}(x_j^-)&=2\nu\left[A_<^*(k')B_<(k)-B_<^*(k')A_<(k)\right],\nonumber\\
 J_{k'k}(x_j^+)&=2\nu\left[B_>^*(k')A_>(k)-A_>^*(k')B_>(k)\right].
\end{align}
The Coulomb corrections do not change these limiting forms.
Thus continuity requires
\begin{equation}
 A_<^*(k')B_<(k)-B_<^*(k')A_<(k)
 =B_>^*(k')A_>(k)-A_>^*(k')B_>(k),
 \label{eq:general-matching}
\end{equation}
which is Eq.~(26) of Ref.~\cite{CamaraDaSilvaLattice} in the present
notation. More general matching conditions can mix the two local
behaviors. Here we choose the simplest real, energy-independent rule,
which does not mix them and introduces neither a length scale nor an
additional matching parameter:
\begin{equation}
 A_<(k)=-A_>(k),\qquad B_<(k)=B_>(k),\qquad
 \eta=\begin{pmatrix}-1&0\\0&1\end{pmatrix}.
 \label{eq:lattice-matching}
\end{equation}
It satisfies Eq.~\eqref{eq:general-matching} and is imposed at every
singularity.

The relative signs have a direct interpretation in the free limit
$g=b=0$, where the powers become $|x-x_j|$ and a constant. The matching
then joins the first mode into $x-x_j$ and continues the second one
unchanged, recovering the usual locally odd and even solutions and
continuity of both $\psi$ and $\psi'$. For nonzero $b$, this interpretation
applies to the leading behaviors; the Coulomb corrections do not have
definite parity across the singularity.

The matching in Eq.~\eqref{eq:lattice-matching} fixes the relation
between opposite sides of a singularity. It does not prescribe a
fixed ratio $B/A$ for a Bloch state. That ratio is determined globally
by propagation through the cell and by Bloch's condition, and therefore
depends on the energy and the Bloch momentum \cite{CamaraDaSilvaLattice}.

\subsection{Transfer matrix and dispersion relation}
For $\nu\neq1/2$, the connection matrix $M(k)$ propagates the coefficients from the
right of $x_j$ to the left of $x_{j+1}$. Applying
Eq.~\eqref{eq:lattice-matching} at $x_{j+1}$ then gives
\begin{equation}
 \begin{pmatrix}A_{j+1,>}(k)/(2i\alpha)^\nu\\
 B_{j+1,>}(k)/(2i\alpha)^{-\nu}\end{pmatrix}
 =T(k)\begin{pmatrix}A_{j,>}(k)/(2i\alpha)^\nu\\
 B_{j,>}(k)/(2i\alpha)^{-\nu}\end{pmatrix},
\end{equation}
where
\begin{equation}
 T(k)=\eta M(k)=
 \begin{pmatrix}-M_{11}(k)&-M_{12}(k)\\M_{21}(k)&M_{22}(k)\end{pmatrix},
 \qquad \det T=1.
 \label{eq:lattice-transfer}
\end{equation}
Bloch's condition,
\begin{equation}
 \psi_{k,q}(x+\ell)=e^{iq\ell}\psi_{k,q}(x),\qquad
 -\alpha\leq q\leq\alpha,
\end{equation}
requires $e^{iq\ell}$ to be an eigenvalue of $T(k)$. Hence
\begin{equation}
 \cos(q\ell)=\frac12\operatorname{tr}T(k)
 =\frac{M_{22}(k)-M_{11}(k)}2.
\end{equation}
Substituting the exact diagonal entries of $M$, the terms proportional
to $\sin(\pi u)\sinh(\pi b/u)$ cancel. The dispersion relation is
\begin{equation}
 \cos\left(\frac{\pi q}{\alpha}\right)
 =\frac{\cos(\pi u)\cosh(\pi b/u)}{\sin(\pi\nu)}.
 \label{eq:lattice-dispersion}
\end{equation}
Appendix~\ref{app:g0} shows explicitly that this relation extends to
$\nu=1/2$. For real energy, its right-hand side is real. The allowed
bands satisfy
\begin{equation}
 \left|\frac{\cos(\pi u)\cosh(\pi b/u)}{\sin(\pi\nu)}\right|
 \leq1,
 \label{eq:lattice-allowed}
\end{equation}
with equality at their edges. In the gaps the Bloch multipliers leave
the unit circle.

For $b>0$, the parametrization introduced in Eq.~\eqref{eq:parameters}
is valid for both signs of the energy:
\begin{equation}
 E(u)=\frac{\hbar^2\alpha^2}{2m}
             \left(u^2-\frac{b^2}{u^2}\right),\qquad u>0.
\end{equation}
Since $E(u)$ is strictly increasing, the allowed intervals in $u$ map
directly to energy bands. At every $u=n+1/2$, $n=0,1,\ldots$, the
right-hand side of Eq.~\eqref{eq:lattice-dispersion} vanishes. Thus each such point lies inside an allowed
band for every $0<\nu<1$.

The limit $b=0$ is obtained directly from
Eq.~\eqref{eq:lattice-dispersion}. For $E\geq0$, it gives
$\cos(\pi q/\alpha)=\cos(\pi k/\alpha)/\sin(\pi\nu)$;
negative energies lie in a gap. Taking also $\nu\to1/2$ recovers
the folded free-particle dispersion.

\subsection{Band widths and gaps}
\label{sec:band-widths}
Let $u_n^-$ and $u_n^+$ denote the lower and upper edges of the band
containing $u=n+1/2$, with $n=0,1,\ldots$. Equation~\eqref{eq:lattice-allowed}
gives the implicit edge equations
\begin{equation}
 u_n^\pm=n+\frac12\pm\frac1\pi
 \arcsin\left[\frac{\sin(\pi\nu)}{\cosh(\pi b/u_n^\pm)}\right].
 \label{eq:width-edges}
\end{equation}
There is one lower edge in $(n,n+1/2)$ and one upper edge in $(n+1/2,n+1)$. Since $E(u)$ is increasing, the width $W_n$ of this band and the gap $G_n$ immediately above it are
\begin{align}
 \frac{W_n}{E_\alpha}
 &=[(u_n^+)^2-(u_n^-)^2]
 \left[1+\frac{b^2}{(u_n^+u_n^-)^2}\right],\nonumber\\
 \frac{G_n}{E_\alpha}
 &=[(u_{n+1}^-)^2-(u_n^+)^2]
 \left[1+\frac{b^2}{(u_{n+1}^-u_n^+)^2}\right],
 \qquad E_\alpha=\frac{\hbar^2\alpha^2}{2m}.
 \label{eq:band-gap-widths}
\end{align}
These expressions, together with Eq.~\eqref{eq:width-edges}, determine the widths for arbitrary $b$. They are invariant under $\nu\to1-\nu$, as is the dispersion relation. For \(b=0\), the band widths and gaps reduce to those of the trigonometric Pöschl–Teller lattice studied in Ref. \cite{CamaraDaSilvaLattice}.

For fixed $b$, the high-band expansion follows by using $\cosh(\pi b/u)=1+\pi^2b^2/(2u^2)+O(u^{-4})$ in the edge equation. In particular,
\begin{align}
 u_n^-&=n+|1/2-\nu|+
 \frac{\pi b^2\cot(\pi|1/2-\nu|)}{2(n+|1/2-\nu|)^2}+O(n^{-4}),\nonumber\\
 u_n^+&=n+1-|1/2-\nu|-
 \frac{\pi b^2\cot(\pi|1/2-\nu|)}{2(n+1-|1/2-\nu|)^2}+O(n^{-4}).
\end{align}
Substitution in Eq.~\eqref{eq:band-gap-widths} gives
\begin{align}
 \frac{W_n}{E_\alpha}
 &=(1-|1-2\nu|)(2n+1)
 -\frac{2\pi b^2\cot(\pi|1/2-\nu|)}{n+1/2}+O(n^{-3}),\nonumber\\
 \frac{G_n}{E_\alpha}
 &=|2-4\nu|(n+1)
 +\frac{2\pi b^2\cot(\pi|1/2-\nu|)}{n+1}+O(n^{-3}).
 \label{eq:widths-high-bands}
\end{align}
The first Coulomb correction therefore narrows the bands and widens the gaps. Its relative contribution vanishes at large $n$, leaving the inverse-square coupling to determine the leading widths.
This expansion is not uniform as $\nu\to1/2$.

At fixed $n$ and large $b$, both edges instead approach $n+1/2$. Expanding Eq.~\eqref{eq:width-edges} about this point yields
\begin{equation}
 \frac{W_n}{E_\alpha}\approx
 \frac{4\sin(\pi\nu)}{\pi}
 \left(n+\frac12+\frac{b^2}{(n+1/2)^3}\right)
 \operatorname{sech}\left(\frac{\pi b}{n+1/2}\right).
 \label{eq:widths-large-b}
\end{equation}
The width is exponentially small in this limit. In particular, low bands can be very narrow without disappearing.


\subsection{Band edges and effective masses}
\label{sec:effective-masses}
Let $(u_e,q_e)$ denote a band edge and define
\begin{equation}
 \mathcal D(u)=\frac{\cos(\pi u)\cosh(\pi b/u)}{\sin(\pi\nu)},
 \qquad \eta_e=\mathcal D(u_e)=\cos(q_e\ell)=\pm1.
\end{equation}
At a simple edge, $\mathcal D'(u_e)\neq0$. Expanding the dispersion
relation $\cos(q\ell)=\mathcal D(u)$ about the edge gives
\begin{equation}
 E(q)=E_e+\frac{\hbar^2}{2m^*(u_e)}(q-q_e)^2
       +O\bigl((q-q_e)^4\bigr),
 \label{eq:edge-quadratic}
\end{equation}
where $E_e=E(u_e)$ and the signed effective mass is
\begin{equation}
 \frac{m^*(u_e)}{m}
 =-\frac{\eta_e\,\mathcal D'(u_e)}
         {\pi^2(u_e+b^2/u_e^3)}.
 \label{eq:effective-mass}
\end{equation}
At the bottom of a band, the curvature is positive and the electron
effective mass is $m_{\mathrm e}^*=m^*(u_e)$. At the top, the
curvature is negative and the hole effective mass is
$m_{\mathrm h}^*=-m^*(u_e)$. In successive bands, the minimum
alternates between $q=0$ and $q=\alpha$, with the maximum at the
opposite edge.

\begin{figure}[!htbp]
\centering\includegraphics[width=.98\linewidth]{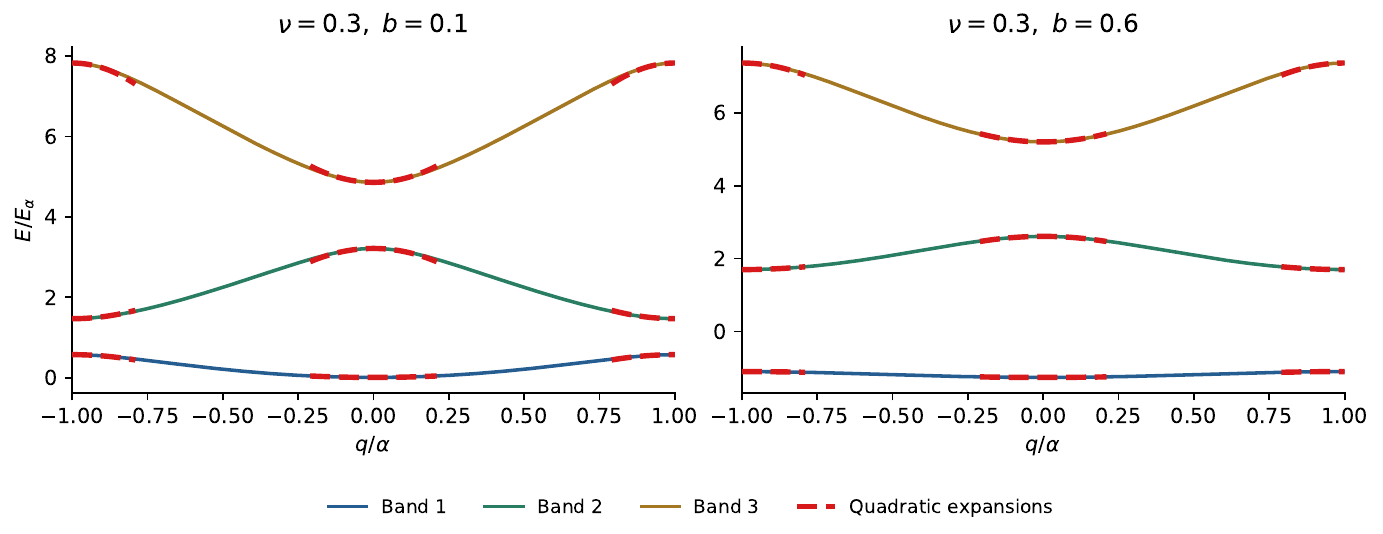}
\caption{The first three exact Bloch bands for $\nu=0.3$, at $b=0.1$
(left) and $b=0.6$ (right). Thick red dashed segments show the quadratic
expansions at every minimum and maximum, restricted to
$|q-q_e|\leq0.21\alpha$. 
Energies are measured in $E_\alpha=\hbar^2\alpha^2/(2m)$.}
\label{fig:band-masses}
\end{figure}
Figure~\ref{fig:band-masses} compares the exact dispersion with these
local parabolas. For both Coulomb couplings shown, the quadratic
approximation describes the dispersion near the band edges.

Table~\ref{tab:effective-masses} lists the edge positions and signed masses
for the first three bands at $\nu=0.3$ and $b=0.4$.

\begin{table}[!htbp]
\centering
\caption{Edges and curvature masses of the first three bands at
$\nu=0.3$, $b=0.4$. The roots $u_e$ are obtained from
$\mathcal D(u_e)=\pm1$ and the masses from
Eq.~\eqref{eq:effective-mass}. The hole masses equal the absolute values
of the negative entries.}
\label{tab:effective-masses}
\begin{tabular}{clrrrr}
\hline
 Band & Edge & $q_e/\alpha$ & $u_e$ & $E_e/E_\alpha$ & $m^*(u_e)/m$\\
\hline
1 & Bottom & 0 & 0.465466 & -0.521831 & 1.707971 \\
 1 & Top & 1 & 0.552701 & -0.218289 & -0.997694 \\
 2 & Bottom & 1 & 1.317374 & 1.643279 & 0.394238 \\
 2 & Top & 0 & 1.717843 & 2.896767 & -0.207613 \\
 3 & Bottom & 0 & 2.254181 & 5.049845 & 0.149666 \\
 3 & Top & 1 & 2.761372 & 7.604192 & -0.104453 \\
\hline
\end{tabular}
\end{table}
For $b=0$, we recover, as expected, the effective mass at the band
edges of the trigonometric P\"oschl--Teller lattice
\cite{CamaraDaSilvaLattice},
\begin{equation}
 \frac{m^*(u_e)}{m}
 =\frac{\eta_e\sin(\pi u_e)}{\pi u_e\sin(\pi\nu)}.
\end{equation}


\subsection{Bloch eigenfunctions and probability density}
\label{sec:probability-density}
For $\nu\neq1/2$, the coefficient ratio in a Bloch state follows from the eigenvector of
$T(k)$ with eigenvalue $\lambda_q=e^{iq\ell}$. The first row gives
\begin{equation}
 \frac{B_0(k,q)}{A_0(k,q)}
 =-(2i\alpha)^{-2\nu}
   \frac{e^{iq\ell}+M_{11}(k)}{M_{12}(k)}.
 \label{eq:bloch-ratio}
\end{equation}
This expression assumes $A_0M_{12}\neq0$; if $A_0=0$, the state
is written with $B_0$ as its reference amplitude. The energy and Bloch
momentum are related by Eq.~\eqref{eq:lattice-dispersion}.

For real energies and $b>0$, the explicit matrix entries simplify
Eq.~\eqref{eq:bloch-ratio} to
\begin{eqnarray}
 \alpha^{2\nu}\frac{B_0(k,q)}{A_0(k,q)}
 =\frac{\nu\,\Gamma(\tfrac12-\nu+u)
                     \Gamma(\tfrac12-\nu-u)
                     |\Gamma(\tfrac12-\nu+ib/u)|^2}
 {\pi\,2^{2\nu}\Gamma(1-2\nu)^2}\left[i\sin(q\ell)
 -\frac{\sin(\pi u)\sinh(\pi b/u)}{\cos(\pi\nu)}\right].
 \label{eq:bloch-ratio-explicit}
\end{eqnarray}
The phases from $(2i\alpha)^{-2\nu}$ and $M_{12}$ cancel. The ratio
is real at a band edge and generally complex in its interior. It is
determined by the lattice state, rather than being an independent
boundary parameter. The finite coefficient ratio and densities at
$\nu=1/2$ are obtained separately in Appendix~\ref{app:g0}.

The density
\begin{equation}
 \rho_{k,q}(x)=|\psi_{k,q}(x)|^2,\qquad
 \rho_{k,q}(x+\ell)=\rho_{k,q}(x),\qquad
 \int_0^\ell\rho_{k,q}(x)\,dx=1
\end{equation}
is periodic even though the wavefunction carries the Bloch phase.
Time reversal gives the same density at $q$ and $-q$.

\begin{figure}[!htbp]
\centering\includegraphics[width=.98\linewidth]{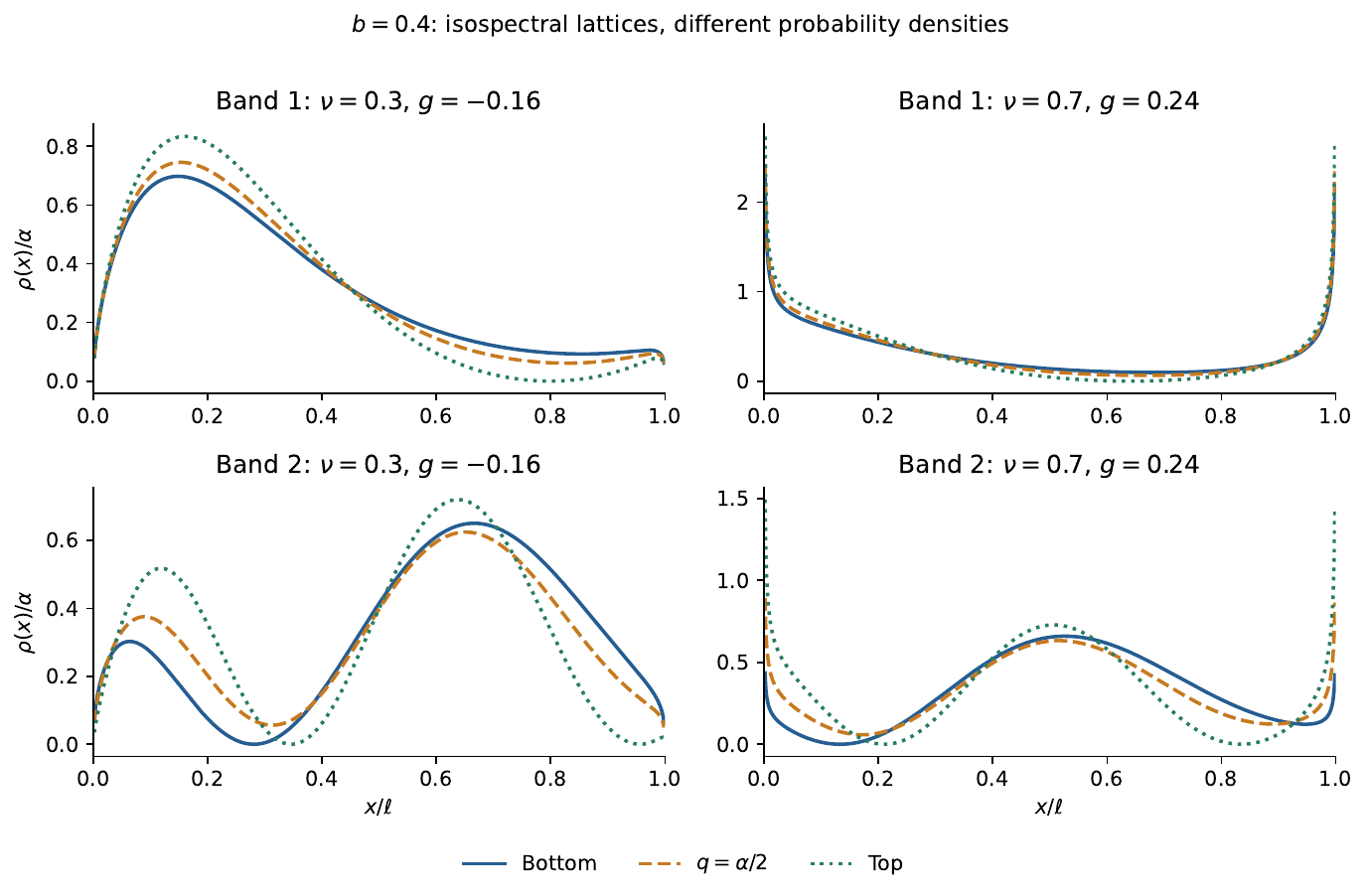}
\caption{Cell-normalized probability densities in the first two bands
(upper and lower rows), for $b=0.4$. The left and right columns show
$\nu=0.3$ and $\nu=0.7$, respectively. Each panel compares the bottom,
the state at $q=\alpha/2$, and the top of the band. The bottom is at
$q=0$ in the first band and at $q=\alpha$ in the second; the top is at
the opposite edge. The lattices are isospectral, but their densities
differ.}
\label{fig:probability-density}
\end{figure}
The dispersion relation is invariant under $\nu\mapsto1-\nu$,
because it depends on $\nu$ only through $\sin(\pi\nu)$. These
lattices therefore have identical band edges and effective masses.
Their local exponents and coefficient ratios, however, are different,
so the spectral symmetry does not extend to their probability
densities, as in the P\"oschl--Teller lattice
\cite{CamaraDaSilvaLattice}.

Figure~\ref{fig:probability-density} compares $\nu=0.3$ ($g=-0.16$)
and $\nu=0.7$ ($g=0.24$) at the same $b=0.4$. Corresponding curves
have the same energy and Bloch momentum. For $\nu=0.3$, the first-band
density is concentrated mainly on the left half of the cell, where the
Coulomb term is attractive, with a broad maximum in the interior.
For $\nu=0.7$, the first-band density instead rises
at both singularities and is depleted in the interior. Thus changing
an attractive inverse-square coupling into a repulsive one does not
remove the endpoint enhancement: the transmitting boundary
conditions retain the irregular local mode.

In the second band, the density develops two broad lobes for
$\nu=0.3$, separated by an interior minimum,
with the larger lobe on the right half of the cell. For $\nu=0.7$,
a broad interior maximum develops between two minima, while the
endpoint peaks remain. Within each band the heights of the lobes and the positions of the minima vary with the Bloch momentum. 

For $B_0\neq0$, the leading density near either singularity is
\begin{equation}
 \rho_{k,q}(x)\approx |B_j(k,q)|^2|x-x_j|^{1-2\nu},
 \label{eq:density-endpoint}
\end{equation}
where $B_j$ denotes the coefficient of the normalized state on the
chosen side. The density vanishes for $0<\nu<1/2$, and diverges
integrably for $1/2<\nu<1$. The latter behavior is compatible with
normalizability and with the transmitting domain even when $g>0$.
If the irregular coefficient vanishes, the regular power
$|x-x_j|^{1+2\nu}$ takes its place. The Coulomb term affects the
subleading corrections on the two sides.

\section{Conclusions}
We have studied the trigonometric Rosen--Morse potential as an isolated
singular well and as an exactly solvable periodic lattice. The model
combines inverse-square singularities of equal strength with Coulomb
contributions of opposite signs at the two ends of each cell. The
connection between the endpoint coefficients allows us to determine
both the renormalized isolated-cell spectrum and the Bloch bands.

For $g\geq-1/4$, the SUSY-compatible condition $B_0=B_1=0$ selects
the known confined spectrum, with Coulomb-dominated low modes for
sufficiently large $b$ and quadratic growth at large mode number.
For $g<-1/4$, renormalization introduces two independent boundary
phases. Deep negative-energy levels form two asymptotic sequences
with the same scaling factor
$\kappa_{l,n+1}/\kappa_{l,n}\to e^{\pi/|\nu|}$ and opposite first
Coulomb corrections. The scaling holds within each endpoint family;
the interlaced spectrum need not form a single geometric sequence.
For a finite core radius $R$, these conclusions have an effective
interpretation within the window $\alpha/\kappa\ll1$,
$\alpha b/\kappa\ll1$ and $\kappa R\ll1$. The infinite descent belongs
to the ideal zero-cutoff Hamiltonian; a core that makes the complete
potential bounded from below restores a ground state.

On the line, transmitting self-adjoint matching conditions couple
neighboring cells while retaining both square-integrable local modes
for $0<\nu<1$, including the logarithmic case $\nu=1/2$. This gives
an exact lattice of inverse-square and Coulomb-type interactions.
The Bloch dispersion determines the band widths, gaps and electron
and hole effective masses. For fixed coupling away from $\nu=1/2$,
both the band widths and gaps grow linearly with the band index at
high energy. The first Coulomb correction narrows the bands and
widens the gaps, whereas at fixed band index and large $b$ the bands
become exponentially narrow. The quadratic expansions reproduce
the dispersion near each separated band edge. The symmetry
$\nu\mapsto1-\nu$ preserves the bands, gaps and effective masses,
but changes the probability densities. Lattices with identical spectra
can therefore have different spatial distributions of states within
each cell.

Together with the previous analyses of singular
P\"oschl--Teller systems
\cite{CAMARADASILVA2024169549z,CamaraDaSilvaTPT,CamaraDaSilvaLattice},
these results extend the renormalization analysis to isolated and
periodic potentials with inverse-square singularities. The boundary
conditions describe renormalization or transmission, while the
connection matrix relates the solutions at successive singularities.
For potentials that are not exactly solvable, this relation can be
obtained numerically. It would also be interesting to study the
hyperbolic counterpart of the present potential, combining
renormalized boundary conditions with the analysis of scattering
amplitudes and their bound, virtual and resonance poles.

The same framework also allows an alternative Rosen-Morse lattice obtained by reversing the sign of \(b\) between neighboring cells. Its local Coulomb terms behave as \(1/|x-x_j|\), with alternating attractive and repulsive centers, each having the same character on both sides of the singularity. With suitable matching conditions, this extension provides a setting to investigate gap closure and the crossover between quadratic and linear dispersion.

A further direction concerns the supersymmetric deformations of the Rosen--Morse potential. First- and second-order SUSY transformations generate new solvable partners, including isospectral families whose endpoint inverse-square singularities are preserved \cite{FernandezReyes2021}. Their additional deformation parameters offer a setting in which to apply the present boundary analysis to a broader class of singular wells. For periodic extensions, one must determine the endpoint couplings and the action of the intertwining operators on the transmitting domains. Periodicity must also be established for the transformed potential, or imposed by repeating a cell with compatible matching data. Isospectrality of the isolated wells alone does not establish equality of their Bloch bands. Comparing the resulting gaps, masses and densities would clarify which properties survive the transformation.

Finally, the matching parameters themselves can be varied across
the lattice. Periodic modulation defines supercells, while a local
change can introduce a defect. Darboux transformations of periodic
potentials already provide examples of defects supporting states
inside spectral gaps \cite{FernandezEtAl2002}. Combining such
constructions with renormalized inverse-square singularities is
a further prospect for studying how local boundary data control
extended and localized states within the same quantum system.

\ack{We thank G. M. Sotkov for a critical reading of the manuscript.}

\appendix
\section{The periodic lattice at $g=0$}
\label{app:g0}
\renewcommand{\theequation}{\thesection.\arabic{equation}}
\renewcommand{\theHequation}{\thesection.\arabic{equation}}
\setcounter{equation}{0}
At $g=0$, the solution with a nonzero endpoint value acquires a
logarithmic correction, and the coefficients used in
Sec.~\ref{sec:lattice} become singular. To describe this case, we
introduce finite endpoint coefficients near $\nu=1/2$ and take the
limit after deriving the connection and matching relations.
The reference scale in the logarithm remains the same $\alpha$ as in
Eq.~\eqref{eq:bc-coulomb-log}.

\subsection{Finite endpoint coefficients}
Write
\begin{equation}
 \nu=\tfrac12+\varepsilon,\qquad
 A_l=\widehat A_l-(-1)^l\frac{b\alpha^{1+2\varepsilon}}{\varepsilon}
 \widehat B_l,\qquad B_l=\widehat B_l,\quad l=0,1.
 \label{eq:app-coefficients}
\end{equation}
Initially $0<|\varepsilon|<1/2$; here $\varepsilon$ measures the
departure from $\nu=1/2$. The symbols without hats retain
their definitions from Sec.~2; the hatted coefficients are finite
in the logarithmic limit. Equation~\eqref{eq:general} becomes
\begin{align}
 \psi_k(x)={}&\left[\widehat A_0(k)
 -\frac{b\alpha^{1+2\varepsilon}}{\varepsilon}\widehat B_0(k)\right]
 \left(\frac{z}{2i\alpha}\right)^{1+\varepsilon}
 \mathcal F(\tfrac12+\varepsilon;-u;z)\nonumber\\
 &+\widehat B_0(k)\left(\frac{z}{2i\alpha}\right)^{-\varepsilon}
 \mathcal F(-\tfrac12-\varepsilon;u;z).
 \label{eq:app-solution}
\end{align}
The functions $\mathcal F$ and the path $z(x)$ are unchanged. The
subtraction acts on the complete solutions, whose pole residues are
proportional; it does not truncate the irregular solution.

Let $\xi=x$ near $x_0=0$ and $\xi=\pi/\alpha-x$ near
$x_1=\pi/\alpha$. The two endpoint expansions, including the opposite
Coulomb corrections, are
\begin{align}
 \psi_k(x)\approx{}&
 \left[\widehat A_l-(-1)^l
 \frac{b\alpha^{1+2\varepsilon}}{\varepsilon}\widehat B_l\right]
 \xi^{1+\varepsilon}
 \left[1-\frac{(-1)^l\alpha b}{1+\varepsilon}\xi+O(\xi^2)\right]
 \nonumber\\
 &+\widehat B_l\xi^{-\varepsilon}
 \left[1+\frac{(-1)^l\alpha b}{\varepsilon}\xi+O(\xi^2)\right],
 \qquad l=0,1.
 \label{eq:app-endpoints}
\end{align}
To take the limit $\varepsilon\to0$, we first expand, at fixed
$\xi>0$,
\begin{align*}
 \xi^{\pm\varepsilon}
 &=1\pm\varepsilon\ln\xi+O(\varepsilon^2),\\
 \alpha^{2\varepsilon}
 &=1+2\varepsilon\ln\alpha+O(\varepsilon^2),\\
 \frac{1}{1+\varepsilon}
 &=1-\varepsilon+O(\varepsilon^2).
\end{align*}
Substituting these expansions into Eq.~\eqref{eq:app-endpoints},
the terms proportional to $1/\varepsilon$ cancel between the two solutions. Collecting the remaining finite terms, the limiting solution has
the asymptotic form
\begin{align*}
 \psi_k(x)={}&
 \widehat A_l(k)\,\xi
 \left[1-(-1)^l\alpha b\,\xi+O(\xi^2)\right]\\
 &+\widehat B_l(k)
 \left[
 1-2(-1)^l\alpha b\,\xi\ln(\alpha\xi)
 +O\!\left(\xi^2|\ln(\alpha\xi)|\right)
 \right],
 \qquad l=0,1.
\end{align*}
This reproduces Eq.~\eqref{eq:bc-coulomb-log}, with its coefficients
identified with $\widehat A_l$ and $\widehat B_l$. The higher-order
terms of the two original series leave finite regular and
logarithmic corrections, included in the remainders above.

The changes in Eq.~\eqref{eq:app-coefficients} do not change the
boundary form. On each side of a singularity their extra terms are
the same real multiple of $\widehat B^*(k')\widehat B(k)$ in the two
products of $J_{k'k}$ and cancel. Consequently,
\begin{align}
 J_{k'k}(x_j^-)&=2\nu\left[
 \widehat A_<^*(k')\widehat B_<(k)
 -\widehat B_<^*(k')\widehat A_<(k)\right],\nonumber\\
 J_{k'k}(x_j^+)&=2\nu\left[
 \widehat B_>^*(k')\widehat A_>(k)
 -\widehat A_>^*(k')\widehat B_>(k)\right].
 \label{eq:app-current}
\end{align}
These limits remain valid at $\varepsilon=0$, with $2\nu=1$:
the logarithmic terms likewise cancel within each one-sided boundary
form. Because the coefficient shifts have opposite signs across a
site, the original matching in Eq.~\eqref{eq:lattice-matching} is
also equivalent to
\begin{equation}
 \widehat A_<=-\widehat A_>,\qquad
 \widehat B_<=\widehat B_>.
 \label{eq:app-matching}
\end{equation}
At $g=0$ it joins continuous wavefunctions and equal finite parts of
the derivatives; the ordinary derivatives have identical logarithmic
divergences. Thus the limiting procedure preserves the transmitting
domain, rather than selecting a different self-adjoint realization.

\subsection{Connection matrix and Bloch dispersion}
The connection already calculated in Sec.~2 gives
\begin{equation}
 \begin{pmatrix}
 \displaystyle\frac{\widehat A_1+
 \frac{b\alpha^{1+2\varepsilon}}{\varepsilon}\widehat B_1}
 {(2i\alpha)^\nu}\\[5pt]
 \displaystyle\frac{\widehat B_1}{(2i\alpha)^{-\nu}}
 \end{pmatrix}
 =M(k)
 \begin{pmatrix}
 \displaystyle\frac{\widehat A_0-
 \frac{b\alpha^{1+2\varepsilon}}{\varepsilon}\widehat B_0}
 {(2i\alpha)^\nu}\\[5pt]
 \displaystyle\frac{\widehat B_0}{(2i\alpha)^{-\nu}}
 \end{pmatrix}.
 \label{eq:app-shifted-connection}
\end{equation}
Solving for the finite coefficients at the second endpoint defines
\begin{equation}
 \begin{pmatrix}\widehat A_1/(2i\alpha)^\nu\\
 \widehat B_1/(2i\alpha)^{-\nu}\end{pmatrix}
 =\widehat M(k)
 \begin{pmatrix}\widehat A_0/(2i\alpha)^\nu\\
 \widehat B_0/(2i\alpha)^{-\nu}\end{pmatrix},
 \label{eq:app-hat-connection}
\end{equation}
where
\begin{align}
 \widehat M_{11}&=M_{11}
 -\frac{b(2i)^{-1-2\varepsilon}}{\varepsilon}M_{21},\nonumber\\
 \widehat M_{22}&=M_{22}
 -\frac{b(2i)^{-1-2\varepsilon}}{\varepsilon}M_{21},\nonumber\\
 \widehat M_{12}&=M_{12}
 -\frac{b(2i)^{-1-2\varepsilon}}{\varepsilon}(M_{11}+M_{22})
 +\frac{b^2(2i)^{-2-4\varepsilon}}{\varepsilon^2}M_{21},\nonumber\\
 \widehat M_{21}&=M_{21},\qquad\det\widehat M=-1.
 \label{eq:app-hat-matrix}
\end{align}
All combinations are formed before taking $\varepsilon\to0$.
Combining the connection with Eq.~\eqref{eq:app-matching} gives the
cell transfer matrix. Its determinant is one, and the two diagonal
shifts cancel from its trace:
\begin{align}
 \widehat T&=\eta\widehat M
 =\begin{pmatrix}-\widehat M_{11}&-\widehat M_{12}\\
 \widehat M_{21}&\widehat M_{22}\end{pmatrix},
 \qquad\det\widehat T=1,\nonumber\\
 \operatorname{tr}\widehat T
 &=\widehat M_{22}-\widehat M_{11}
 =M_{22}-M_{11}
 =2\frac{\cos(\pi u)\cosh(\pi b/u)}{\cos(\pi\varepsilon)}.
 \label{eq:app-trace}
\end{align}
Therefore the Bloch relation at nonzero $\varepsilon$ is exactly
Eq.~\eqref{eq:lattice-dispersion} with
$\nu=1/2+\varepsilon$. Taking the limit gives
\begin{equation}
 \cos(q\ell)=\mathcal D_0(u),\qquad
 \mathcal D_0(u)=\cos(\pi u)\cosh(\pi b/u),\qquad
 \frac{E}{E_\alpha}=u^2-\frac{b^2}{u^2}.
 \label{eq:app-dispersion}
\end{equation}
This is the main-text dispersion evaluated at $\nu=1/2$, now derived
in a basis with a finite logarithmic limit.

To obtain the limiting matrix and its eigenvectors explicitly, write
\begin{equation}
 (2i)^{-2\nu}M_{21}
 =\frac{\pi\,2^{-2\nu}\Gamma(1+2\nu)^2}
 {\nu\,\Gamma(\tfrac12+\nu+u)\Gamma(\tfrac12+\nu-u)
 |\Gamma(\tfrac12+\nu+ib/u)|^2}
 =\mathcal C_0(u)\left[1+\varepsilon\mathcal L(u)+O(\varepsilon^2)\right],
 \label{eq:app-C-expansion}
\end{equation}
where the Gamma reflection and recurrence identities give
\begin{align}
 \mathcal C_0(u)&=\frac{\sin(\pi u)\sinh(\pi b/u)}{\pi b},\nonumber\\
 \mathcal L(u)&=2-4\gamma_{\mathrm E}-2\ln2
 -\psi(1+u)-\psi(1-u)
 -2\operatorname{Re}\psi(1+ib/u).
 \label{eq:app-finite-functions}
\end{align}
Here $\psi$ is the digamma function and $\gamma_{\mathrm E}$ is Euler's
constant. The pole in the diagonal entries of $M$ satisfies
$\sin(\pi u)\sinh(\pi b/u)/\cos(\pi\nu)
=-b\mathcal C_0(u)/\varepsilon+O(\varepsilon)$.
It cancels against the shifted $M_{21}$ term, leaving the finite matrix
\begin{equation}
 \widehat M^{(0)}\equiv\lim_{\varepsilon\to0}\widehat M
 =\begin{pmatrix}
 -\mathcal D_0-b\mathcal C_0\mathcal L&
 \displaystyle\frac{1+b^2\mathcal C_0^2\mathcal L^2-\mathcal D_0^2}
 {2i\mathcal C_0}\\[5pt]
 2i\mathcal C_0&\mathcal D_0-b\mathcal C_0\mathcal L
 \end{pmatrix}.
 \label{eq:app-limiting-matrix}
\end{equation}
The upper-right entry follows from
$\det\widehat M^{(0)}=-1$. 
The corresponding transfer matrix is
$\widehat T^{(0)}=\eta\widehat M^{(0)}$, with trace
$2\mathcal D_0(u)$.

\subsection{Coefficient ratio and probability densities}
The second row of the Bloch eigenvector equation gives a finite
coefficient ratio entirely in terms of the energy parameter:
\begin{eqnarray}
 \alpha\frac{\widehat B_0(k,q)}{\widehat A_0(k,q)}
 \Bigg|_{\varepsilon=0}=\frac{\mathcal C_0(u)}
 {i\sin(q\ell)+b\mathcal C_0(u)\mathcal L(u)}=\left[b\mathcal L(u)
 \pm i\frac{\sqrt{1-\mathcal D_0(u)^2}}{\mathcal C_0(u)}\right]^{-1}.
 \label{eq:app-ratio}
\end{eqnarray}
The signs correspond to opposite Bloch momenta, and the second form
uses Eq.~\eqref{eq:app-dispersion}. 
Inserting Eq.~\eqref{eq:app-ratio} into
Eq.~\eqref{eq:app-solution}, combining its two terms before taking
$\varepsilon\to0$, and normalizing over one cell yields the Bloch
probability density. Opposite momenta give complex-conjugate states
and identical densities.

For $\xi=|x-x_l|$, the local density is
\begin{align}
 \rho_{k,q}(x)={}&|\widehat B_l|^2
 \left[1-4(-1)^l\alpha b\,\xi\ln(\alpha\xi)\right]
 +2\operatorname{Re}(\widehat A_l\widehat B_l^*)\,\xi\nonumber\\
 &+O\!\left(\xi^2\ln^2(\alpha\xi)\right),\qquad l=0,1.
 \label{eq:app-density-log}
\end{align}
The density is finite and continuous across the sites. For
$\widehat B_0\neq0$, its derivative on either side is
$-4\alpha b|\widehat B_0|^2\ln(\alpha\xi)+O(1)$ and diverges
logarithmically.

\begin{figure}[!htbp]
\centering\includegraphics[width=.98\linewidth]{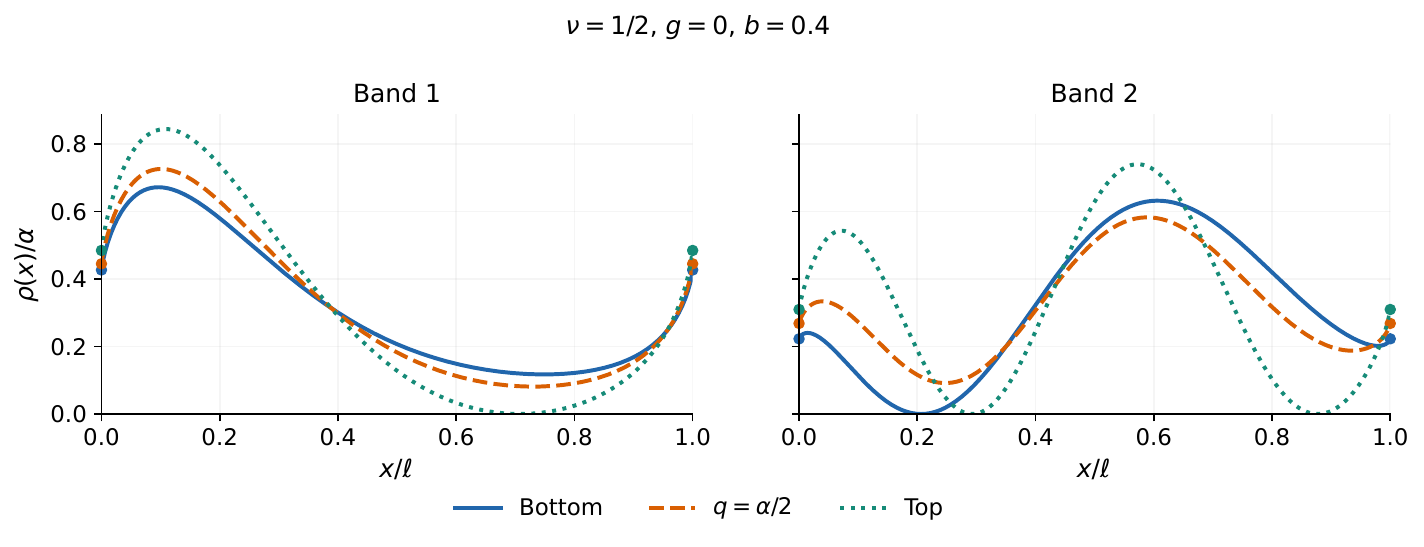}
\caption{Normalized probability densities for $g=0$ ($\nu=1/2$),
$b=0.4$, in the first (left) and second (right) bands. The curves
compare the bottom, the state at $q=\alpha/2$, and the top of each
band. The dots mark the common finite endpoint values
$\rho(0)=\rho(\ell)$.}
\label{fig:probability-density-log}
\end{figure}
Figure~\ref{fig:probability-density-log} shows these densities at the
same $b$ and Bloch momenta as Fig.~\ref{fig:probability-density}.
In the first band, the density is concentrated mainly on the half of
the cell where the Coulomb term is attractive, with a broad maximum
in the interior. In the second band, two interior lobes develop and
the probability is greater on the repulsive half. This redistribution
between the two halves of the cell therefore persists without the
inverse-square term. The site values
are nonzero, in contrast to $0<\nu<1/2$, and finite, in contrast to
$1/2<\nu<1$. Both local modes are retained by the transmitting
domain.

\bibliographystyle{unsrt}
\bibliography{references_review}

@article{Asim_paper,
	author = {Gangopadhyaya, Asim and Panigrahi, Prasanta K. and Sukhatme, Uday P.},
	doi = {10.1088/0305-4470/27/12/032},
	journal = {J. Phys. A: Math. Gen.},
	number = {12},
	pages = {4295--4300},
	title = {Analysis of inverse-square potentials using supersymmetric quantum mechanics},
	volume = {27},
	year = {1994}}

@article{S_Coon,
	author = {Sidney A. Coon and Barry R. Holstein},
	journal = {American Journal of Physics},
	month = {May},
	number = {5},
	pages = {513-519},
	title = {Anomalies in quantum mechanics: The $1/r^2$ potential},
	volume = {70},
	year = {2002}}

@article{CamaraDaSilvaTPT,
	author = {Camara da Silva, Ulysses},
	title = {Renormalization of {Schr{\"o}dinger} equation for potentials with inverse-square singularities: Generalized trigonometric {P{\"o}schl--Teller} model},
	journal = {Journal of Physics A: Mathematical and Theoretical},
	volume = {58},
	number = {50},
	pages = {505201},
	year = {2025},
	doi = {10.1088/1751-8121/ae2446},
	eprint = {2503.12715},
	archiveprefix = {arXiv},
	primaryclass = {quant-ph},
	url = {https://doi.org/10.1088/1751-8121/ae2446}}

@article{Burgess:2017ud,
	author = {Burgess, C. P. and Hayman, Peter and Williams, M. and Zalav{\'a}ri, L{\'a}szl{\'o}},
	da = {2017/04/19},
	doi = {10.1007/JHEP04(2017)106},
	id = {Burgess2017},
	isbn = {1029-8479},
	journal = {Journal of High Energy Physics},
	number = {4},
	pages = {106},
	title = {Point-particle effective field theory I: classical renormalization and the inverse-square potential},
	ty = {JOUR},
	url = {https://doi.org/10.1007/JHEP04(2017)106},
	volume = {2017},
	year = {2017}}

@article{CAMARADASILVA2024169549z,
	author = {Ulysses {Camara da Silva} and Carlos F.S. Pereira and Andre {Alves Lima}},
	doi = {https://doi.org/10.1016/j.aop.2023.169549},
	issn = {0003-4916},
	journal = {Annals of Physics},
	pages = {169549},
	title = {Renormalization group and spectra of the generalized P{\"o}schl--Teller potential},
	url = {https://www.sciencedirect.com/science/article/pii/S0003491623003512},
	volume = {460},
	year = {2024}}

@article{Lima:2019xzg,
	archiveprefix = {arXiv},
	author = {Scursulim, J. V. S. and Camara da Silva, U. and Sotkov, G. M. and Lima, A. A.},
	doi = {10.1103/PhysRevA.101.032105},
	eprint = {1912.13014},
	journal = {Phys. Rev. A},
	number = {3},
	pages = {032105},
	primaryclass = {hep-th},
	title = {{Supersymmetry shielding the scaling symmetry of conformal quantum mechanics}},
	volume = {101},
	year = {2020}}

@article{Bawin:2003dm,
	archiveprefix = {arXiv},
	author = {Bawin, M. and Coon, S. A.},
	doi = {10.1103/PhysRevA.67.042712},
	eprint = {quant-ph/0302199},
	journal = {Phys. Rev. A},
	pages = {042712},
	title = {{The Singular inverse square potential, limit cycles and selfadjoint extensions}},
	volume = {67},
	year = {2003}}

@misc{NIST:DLMF,
	howpublished = {\url{https://dlmf.nist.gov/}, Release 1.1.10 of 2023-06-15},
	key = {{\relax DLMF}},
	note = {F.~W.~J. Olver, A.~B. {Olde Daalhuis}, D.~W. Lozier, B.~I. Schneider, R.~F. Boisvert, C.~W. Clark, B.~R. Miller, B.~V. Saunders, H.~S. Cohl, and M.~A. McClain, eds.},
	title = {{\it NIST Digital Library of Mathematical Functions}},
	url = {https://dlmf.nist.gov/}}

@article{dutt1988,
	author = {Dutt, Ranabir and Khare, Avinash and Sukhatme, Uday P.},
	doi = {10.1119/1.15697},
	journal = {Am. J. Phys.},
	pages = {163--168},
	reportnumber = {UIC-86-12},
	title = {{Supersymmetry, Shape Invariance and Exactly Solvable Potentials}},
	volume = {56},
	year = {1988}}

@article{Gitman,
	archiveprefix = {arXiv},
	author = {Gitman, D. M. and Tyutin, I. V. and Voronov, B. L.},
	doi = {10.1088/1751-8113/43/14/145205},
	eprint = {0903.5277},
	journal = {J. Phys. A: Math. Theor.},
	number = {14},
	pages = {145205},
	primaryclass = {quant-ph},
	title = {{Self-adjoint extensions and spectral analysis in Calogero problem}},
	volume = {43},
	year = {2010}}

@article{renor_orig,
	archiveprefix = {arXiv},
	author = {Beane, S. R. and Bedaque, Paulo F. and Childress, L. and Kryjevski, A. and McGuire, J. and van Kolck, U.},
	doi = {10.1103/PhysRevA.64.042103},
	eprint = {quant-ph/0010073},
	journal = {Phys. Rev. A},
	pages = {042103},
	reportnumber = {NT-UW-00-023, DOE-ER-41132-102-INT-00, RBRC-140, KRL-MAP-271, NT@UW-00-023},
	title = {{Singular potentials and limit cycles}},
	volume = {64},
	year = {2001}}

@article{renor_Group_Limit_Cycle,
	archiveprefix = {arXiv},
	author = {Braaten, Eric and Phillips, Demian},
	doi = {10.1103/PhysRevA.70.052111},
	eprint = {hep-th/0403168},
	journal = {Phys. Rev. A},
	pages = {052111},
	title = {{The Renormalization group limit cycle for the $1/r^2$ potential}},
	volume = {70},
	year = {2004}}

@article{renor_m_w_c,
	author = {Bouaziz, Djamil and Bawin, Michel},
	journal = {Physical Review A},
	number = {2},
	pages = {022113},
	publisher = {APS},
	title = {Singular inverse-square potential: renormalization and self-adjoint extensions for medium to weak coupling},
	volume = {89},
	year = {2014}}

@article{Renor_Inver_Squa,
	archiveprefix = {arXiv},
	author = {Camblong, Horacio E. and Epele, Luis N. and Fanchiotti, Huner and Garcia Canal, Carlos A.},
	doi = {10.1103/PhysRevLett.85.1590},
	eprint = {hep-th/0003014},
	journal = {Phys. Rev. Lett.},
	pages = {1590--1593},
	title = {{Renormalization of the inverse square potential}},
	volume = {85},
	year = {2000}}

@article{Efimov,
  author  = {Efimov, V.},
  title   = {Energy levels arising from resonant two-body forces in a three-body system},
  journal = {Physics Letters B},
  volume  = {33},
  number  = {8},
  pages   = {563--564},
  year    = {1970},
  doi     = {10.1016/0370-2693(70)90349-7}
}

@article{deAlfaro:1976vlx,
	author = {de Alfaro, Vittorio and Fubini, S. and Furlan, G.},
	doi = {10.1007/BF02785666},
	journal = {Nuovo Cim. A},
	pages = {569},
	reportnumber = {CERN-TH-2115},
	title = {{Conformal Invariance in Quantum Mechanics}},
	volume = {34},
	year = {1976}}

@article{GesztesyMacedoStreit1985,
	author = {Gesztesy, Fritz and Macedo, Carlos and Streit, Ludwig},
	title = {An exactly solvable periodic {Schr{\"o}dinger} operator},
	journal = {Journal of Physics A: Mathematical and General},
	volume = {18},
	number = {9},
	pages = {L503--L507},
	year = {1985},
	doi = {10.1088/0305-4470/18/9/003},
	url = {https://doi.org/10.1088/0305-4470/18/9/003}}

@article{GesztesyKirsch1985,
	author = {Gesztesy, Fritz and Kirsch, Werner},
	title = {One-dimensional {Schr{\"o}dinger} operators with interactions singular on a discrete set},
	journal = {Journal f{\"u}r die reine und angewandte Mathematik},
	volume = {362},
	pages = {28--50},
	year = {1985},
	doi = {10.1515/crll.1985.362.28},
	url = {https://doi.org/10.1515/crll.1985.362.28}}

@article{Case1950,
 author={Case, K. M.},title={Singular potentials},
 journal={Phys. Rev.},volume={80},pages={797--806},year={1950},
 doi={10.1103/PhysRev.80.797}}

@article{Frank1971,
 author={Frank, W. M. and Land, D. J. and Spector, R. M.},
 title={Singular potentials},journal={Rev. Mod. Phys.},volume={43},
 pages={36--98},year={1971},doi={10.1103/RevModPhys.43.36}}

@misc{CamaraDaSilvaLattice,
 author={Camara da Silva, Ulysses},
 title={From isolated singular wells to exact {Bloch} bands: the trigonometric {P\"oschl--Teller} lattice},
 year={2026},howpublished={arXiv:2610.16072 [math-ph]},
 url={https://arxiv.org/abs/2610.16072}}

@article{RosenMorse1932,
 author={Rosen, N. and Morse, Philip M.},
 title={On the Vibrations of Polyatomic Molecules},
 journal={Phys. Rev.},volume={42},pages={210--217},year={1932},
 doi={10.1103/PhysRev.42.210}}

@article{Schrodinger1940,
 author={Schr{\"o}dinger, E.},
 title={A Method of Determining Quantum-Mechanical Eigenvalues and Eigenfunctions},
 journal={Proc. R. Irish Acad. A},volume={46},pages={9--16},year={1940},
 url={https://www.jstor.org/stable/20490744}}

@article{FernandezReyes2021,
 author = {Fern{\'a}ndez, David J. and Reyes, Rosa},
 title = {Spectral manipulation of the trigonometric {Rosen--Morse} potential through supersymmetry},
 journal = {Physica Scripta},
 volume = {96}, number = {12}, pages = {125250}, year = {2021},
 doi = {10.1088/1402-4896/ac33f9},
 eprint = {2107.00232}, archivePrefix = {arXiv}, primaryClass = {quant-ph}
}

@article{FernandezEtAl2002,
 author = {Fern{\'a}ndez, David J. and Mielnik, Bogdan and Rosas-Ortiz, Oscar and Samsonov, Boris F.},
 title = {Nonlocal supersymmetric deformations of periodic potentials},
 journal = {Journal of Physics A: Mathematical and General},
 volume = {35}, number = {19}, pages = {4279--4291}, year = {2002},
 doi = {10.1088/0305-4470/35/19/309},
 eprint = {quant-ph/0303051}, archivePrefix = {arXiv}
}
\end{document}